\documentclass[11pt]{elsarticle}
\usepackage[acronym,nomain,nonumberlist]{glossaries}
\usepackage{glossary-mcols}
\makeglossaries

\usepackage{setspace}
\usepackage[top=3.5cm,bottom=3.5cm,left=3.5cm,right=3.5cm]{geometry}

\usepackage{lineno,hyperref}
\modulolinenumbers[5]

\journal{European Journal of Operational Research}

\usepackage{amssymb}
\usepackage{mathtools}
\usepackage{mathrsfs}  

\usepackage{graphicx}  %Required g

\usepackage{amsmath}
\usepackage{graphicx,psfrag,epsf}
\usepackage{enumerate}
\usepackage{natbib}
\usepackage{comment}
\usepackage[official]{eurosym}
\usepackage{eurosym}
\usepackage{url} % not crucial - just used below for the URL 
\usepackage{siunitx}

\usepackage{url}            % simple URL typesetting
\usepackage{booktabs}       % professional-quality tables
\usepackage{subcaption}
\usepackage{amsfonts}       % blackboard math symbols
\usepackage{nicefrac}       % compact symbols for 1/2, etc.
\usepackage{microtype}      % microtypography

\usepackage{amsmath}
\usepackage{mathabx}
\usepackage{amsopn}

\usepackage{color}
\usepackage{caption}

\usepackage{float}
\floatstyle{plaintop}
\restylefloat{table}

\usepackage{amsthm}
\usepackage{bm}

\DeclareMathOperator*{\argmax}{arg\,max}

\DeclareMathOperator*{\mode}{mode}
\newcommand{\Acal}{\mathcal{A}}
\newcommand{\Dcal}{\mathcal{D}}

\newcommand{\Pcal}{\mathcal{P}}
\newcommand{\Ucal}{\mathcal{U}}
\newcommand{\Pbb}{\mathbb{P}}
\newcommand{\opt}{*}
\newcommand{\dd}{\mathop{}\! \mathrm{d}}
\renewcommand{\thefootnote}{\roman{footnote}}
\newcommand\given[1][]{\:#1\vert\:}

\newtheorem{prop}{Proposition}

\usepackage{color,soul}
\usepackage[usestackEOL]{stackengine}[2013-09-11]

\setstackgap{L}{1.4\baselineskip}
\usepackage{tikz}
\usetikzlibrary{arrows,automata,backgrounds,fit,patterns,petri,shadows,shapes}
\pgfdeclarepatternformonly{stripes}
{\pgfpointorigin}{\pgfpoint{0.25cm}{0.25cm}}
{\pgfpoint{0.25cm}{0.25cm}}
{
    \pgfpathmoveto{\pgfpoint{0cm}{0cm}}
    \pgfpathlineto{\pgfpoint{0.25cm}{0.25cm}}
    \pgfpathlineto{\pgfpoint{0.25cm}{0.12cm}}
    \pgfpathlineto{\pgfpoint{0.12cm}{0cm}}
    \pgfpathclose%
    \pgfusepath{fill}
    \pgfpathmoveto{\pgfpoint{0cm}{0.12cm}}
    \pgfpathlineto{\pgfpoint{0cm}{0.25cm}}
    \pgfpathlineto{\pgfpoint{0.12cm}{0.25cm}}
    \pgfpathclose%
    \pgfusepath{fill}
}

\usepackage{tabularx}
\usepackage{booktabs}

\usepackage[boxed, vlined]{algorithm2e}
\SetKwComment{Comment}{$\triangleright$\ }{} % Comments in algos
\SetAlCapSkip{0.5em}
\SetKwInput{Input}{input}
\SetKwInOut{Initialize}{initialize}
\SetEndCharOfAlgoLine{}

\makeglossaries
\newacronym{1}{BNE}{Bayes Nash equilibrium}
\newacronym{2}{ARA}{Adversarial risk analysis}
\newacronym{3}{APS}{Augmented probability simulation}
\newacronym{4}{MC}{Monte Carlo}
\newacronym{5}{NP}{Non-deterministic polynomial time}
\newacronym{6}{AML}{Adversarial machine learning}
\newacronym{7}{SLLN}{Strong law of large numbers}
\newacronym{8}{MCMC}{Markov chain Monte Carlo}
\newacronym{9}{MH}{Metropolis Hastings}
\newacronym{10}{DDoS}{Distributed denial of service}
\newacronym{11}{HMC}{Hamiltonian Monte Carlo}
\newacronym{12}{AI}{Artificial Intelligence}
\newacronym{13}{NE}{Nash equilibrium}
\newacronym{14}{SM}{Supplementary Material}
\newacronym{15}{SSE}{Strong Stackelberg equilibrium}
\biboptions{authoryear}
\begin{document}

\begin{frontmatter}

\title{\textcolor{black}{Augmented Probability Simulation Methods for Sequential Games}}


\author[mymainaddress]{Tahir Ekin}

\author[mysecondaryaddress]{Roi Naveiro\corref{mycorrespondingauthor}}
\ead{roi.naveiro@cunef.edu}
\cortext[mycorrespondingauthor]{Corresponding author}

\author[mythirdaddress]{David Ríos Insua}

\author[myfourthaddress]{Alberto Torres-Barrán}

\address[mymainaddress]{McCoy College of Business, Texas State University.}
\address[mysecondaryaddress]{CUNEF Universidad, Madrid, Spain.}
\address[mythirdaddress]{Institute of Mathematical Sciences (ICMAT), Madrid, Spain.}
\address[myfourthaddress]{Komorebi AI, Madrid, Spain.}

\begin{abstract}
We present a robust framework with computational algorithms to 
support decision makers in sequential games. 
Our framework includes methods to solve games with complete information,
assess the robustness of such solutions and, finally, approximate adversarial risk analysis solutions when lacking complete information. 
Existing simulation based approaches can be inefficient \textcolor{black}{ when dealing with
 large sets of feasible decisions}; the game of interest may not even be 
solvable to the desired precision for continuous decisions. Hence, we provide a novel alternative solution method
based on the use of augmented probability simulation. While the proposed framework \textcolor{black}{conceptually} applies to \textcolor{black}{multi-stage sequential games, the discussion focuses on two-stage sequential defend-attack problems.
}

\end{abstract}

\begin{keyword}
Decision analysis %\sep Sequential decision analysis 
\sep {\color{black}Sequential} %non-cooperative
games \sep Augmented probability simulation \sep Adversarial risk analysis
\end{keyword}

\end{frontmatter}

%\linenumbers

\section{Introduction}
\textcolor{black}{Sequential games refer to decision making environments where 
%correspond to cases where %one 
decision makers choose actions alternately over time. In articular, we shall 
emphasise non-cooperative sequential games, %non-cooperative situations.
%Cooperative sequential games abstracts away from procedures %and concentrates on the possibilities for agreement 
%\citep{brandenburger2007cooperative}, hence, the traditional focus has been on non-cooperative games.}
%\footnote{\textcolor{red}{While sequential games can be classified as either non-cooperative and cooperative games, 
%, which is also the emphasis of this manuscript. }}
%before the others, %choose theirs
%in a cooperative or non-cooperative way.}  
%While there can be cooperation among decision makers, traditional sequential games focus on cases lack of cooperation.
%formalize interdependence among the players. In the non-cooperative theory, a game is a detailed model of all the moves available to the players. By contrast, the cooperative theory abstracts away from this level of detail, and describes only the outcomes that result when the players come together in different combinations
%In particular,} 
%non-cooperative game theory 
covering} %refers to 
conflict situations in which two or more agents make decisions whose payoffs depend on the actions implemented by all of them and, possibly, on some random outcomes. Under complete information %assumptions
about the agents' preferences and beliefs, the analysis is usually done through Nash equilibria \textcolor{black}{(NE)} and related refinements  which provide a prediction of the agents' decisions. %\cite{MO:2011} review these ideas, whereas
On the other hand, games with incomplete information correspond to cases where the agents do not possess full information about their opponents,
\textcolor{black}{ and are traditionally \textcolor{black}{solved with} Bayes Nash equilibrium (BNE) concepts \citep{harsanyi}}. % and refinements}
\textcolor{black}{ \cite{Hargreaves:2004} provide an in-depth critical assessment}.  
\textcolor{black}{The standard common prior hypothesis underlying  BNE \citep{Antos} is relaxed with the adversarial risk analysis (ARA) \citep{Banks:2015} methodology which assesses the prior over agents' types using decision analytic arguments.
In addition, ARA provides}
%Adversarial risk analysis (ARA) \textcolor{black}{ provides an alternative decision analytic approach to games  relaxing, among others, the standard common prior hypothesis underlying  BNE \citep{Antos}. ARA aims to provide 
prescriptive support to one of the decision makers based on a subjective expected utility model encompassing a forecast of the adversaries' decisions. 
\cite{Banks2020} show that the ARA solution algorithmically coincides with the BNE for two-stage sequential games. 
%In addition, ARA offers a decision theoretic way of computing the prior over agents' types.
 % Their (random) optimal actions are predicted taking into account the uncertainty about the adversaries' probabilities and utilities in an expected utility model of their behaviour.
%In contrast with traditional game theoretic approaches, it relaxes the standard common knowledge hypothesis
\textcolor{black}{Further comparisons and discussions of relationships between ARA and traditional game-theoretic concepts can be found in  \cite{Banks:2015,Banks2020}.}

%Building on both methodologies, this paper presents a comprehensive robust %decision support framework for agents in a non-cooperative setups.

Our realm in this paper will be within algorithmic decision \citep{rossi2009algorithmic} and game \citep{Nisan:2007} theories, %with the objective of
proposing scalable methods to approximate solutions for
\textcolor{black}{sequential} 
 games with large (even continuous) decision sets and uncertain outcomes. %\textcolor{red}{These games are seen in various application domains such as security and adversarial machine learning. }%and they are hard to solve.}
%When the games have small number of decision alternatives and/or discrete uncertainty, many solution methods are available including but not limited to numerical enumeration and integration \textcolor{red}{Reference}. 
\textcolor{black}{Discussions concerning the complexity of computing
game-theoretic solutions
in security games may be seen in \cite{korzhyk} and references therein.}
A variety of simulation based approaches could be utilized for cases where analytical solutions are not available  \citep{amaran2016simulation}. % These methods consist of estimating the expected utility functions which are later used in optimization. \textcolor{red}{Is there a well-known book about simulation based methods for games; also maybe we can give names of other simulation based approaches other than MC} 
Among those, Monte Carlo (MC) methods are
straightforward %to use 
and  widely implemented \citep{shao1989monte}. However, they can be inefficient in face of a large number of alternatives,  % and 
 %As an example,
as in counter-terrorism and cybersecurity problems which may involve thousands of possible decisions and large uncertainties about the attackers' goals and resources  \citep{Zhuang:2007}. %This can result in computational challenges especially in cases where model uncertainty dominates, \cite{Rios-Insua:2009}.
In addition, in problems with continuous decision sets and decision dependent uncertainties, MC requires discretizing  decision spaces, with its performance critically depending on the precision of such discretization. Increasing accuracy may require dealing with many more decision alternatives at a prohibitive computational cost.

Procedures  that focus on high-probability high-impact events,
such as importance sampling,
could improve the estimations by reducing their variance. %as part of solution methods for games (\cite{bowling2008strategy}). 
However, an optimization problem would still need to be solved. 
%handle computational challenges due to high dimensions. %
As an alternative, \cite{BMI:1999} introduced
augmented probability simulation (APS) to approximate optimal solutions in decision analytic problems. APS transforms the %decision analysis
problem into a grand simulation %problem 
in the joint space of  
decision and random variables, \textcolor{black}{ constructing 
an auxiliary augmented distribution (from now on, the {\em augmented distribution}) which is proportional to the product of the utility function and 
the original distribution. Then, 
the optimal decision alternative would coincide
with 
the mode 
of the marginal of the augmented 
distribution over the decisions.}
As a consequence, \textcolor{black}{ simulation from the augmented distribution %where 
%not only the outcomes but also the decisions are treated as
%random,
enables undertaking the evaluation and optimization tasks simultaneously.} 

%We should note that 
\textcolor{black}{We focus on comparing APS and MC methods 
in solving games because of their shared broad applicability. }
%Hereafter, our discussion will revolve around the comparison of APS and MC methods because they share the same broad level of applicability. In the discussion, we will briefly present how APS could be enhanced in particular settings where gradient information is available.   } 
{\color{black}  %there could be  
More specific methods could be preferred in certain contexts. %than MC and APS usable to our advantage. 
For instance, when dealing with continuous decision sets,
extensions of simulation optimization methods \citep{nelson2001simple} that incorporate gradient information could be useful, especially in high-dimensional settings
%. As an example, in games with continuous and high dimensional decision sets, available gradient information could be exploited. %A limited version of this was used to solve Stackelberg games
\citep{naveiro2019gradient}. When gradients cannot be directly evaluated, stochastic approximation or response surface methods \citep{fu2015handbook} could be potential alternatives. Nevertheless, these other methods are only applicable for games with continuous decision sets. 

This paper presents} a robust decision support framework with novel APS-based computational algorithms for decision makers in sequential games.
The framework covers cases of both complete and incomplete information,
interlinked through sensitivity analysis.
\textcolor{black}{ To the best of our knowledge,
this is the first sequential use of APS algorithms to solve sequential decision problems with more than one stochastic decision stage and, moreover, with more than one decision maker}. A key advantage of APS in our setting is that, unlike plain MC methods, its complexity does not depend on the cardinality of the decision sets, thus being the preferred approach when dealing with large decision spaces. Indeed, our approach %can naturally deal with continuous decision sets, 
\textcolor{black}{ does not require discretization in continuous decision sets;}
this makes it scalable
in important contexts such as adversarial machine learning (AML) \citep{AMLARA},
\textcolor{black}{ entailing very high dimensional continuous decision spaces and, consequently, 
 hardly solvable using standard methods}. 
 Moreover, APS can be used to sample from a power
 transformation of the distribution of interest, this being 
  more peaked around the mode, thus facilitating identification of the optimal
alternative. \textcolor{black}{ Finally, our approach 
provides sensitivity 
analysis tools at no extra computational cost.}
%In particular, MC and APS based approaches are used to compute solutions for games with complete and incomplete information. 
%\textcolor{red}{Decide if necessary? We illustrate the use of the proposed approach in solution of a cyber security problem.} 

\textcolor{black}{To make the presentation more concise,
we focus on
sequential games in which the supported agent makes a decision first, then observed by another agent who makes
his own decision. These  
correspond to  two-stage
sequential defend-attack \citep{Brown:2006} or, more generally, to Stackelberg  \citep{korzhyk} games. Their importance in the literature
of security games \citep{sinhatambe,Zhuang:2007} and AML 
\citep{AMLARA} inspires our developments. It is also worth mentioning that such games are relevant in numerous other  areas, from its original conception in business competition \citep{stackel} to automated 
driving systems \citep{yu2018human}.}
\textcolor{black}{Although we focus on illustrating  our methods in 
two stage defend-attack games, the framework conceptually extends
to multi-stage sequential games, as %for limited reasons, 
sketched in our discussion and outlined in %some
detail in the supplementary materials, 
including a demonstration for a three stage game.}
%Our final discussion sketches how the presented  concepts can be extended to solve general sequential non-cooperative games}.

\textcolor{black}{The paper proceeds as follows.}
\textcolor{black}{Sections \ref{sec:APS} and \ref{sec:ARA} present the key steps 
of the framework: approximating %subgame perfect 
equilibria under complete information, assessing robustness of such equilibria 
and, finally, if necessary, approximating ARA solutions under incomplete information.} % MC and APS based algorithms are provided to solve each stage.}   
A computational assessment is presented in 
Section \ref{sec:time_comp}, followed by 
a cybersecurity case study in
Section \ref{sec:example2}. 
Section \ref{sec:disc} concludes  
with a discussion. \textcolor{black}{ An Appendix sketches key proofs.
%and detailed explanation of propositions, in addition to the convergence proofs for the proposed algorithms. 
A \textit{Supplementary Materials} (SM) file presents additional results, algorithms, 
and details of the case study.
Code to reproduce the results is available in a GitHub repository \citep{github}.}
%{\color{black}As standard in the probability theory jargon, we shall refer to https://www.overleaf.com/project/61ed6ca805a5c3a11f8d9023events happening almost surely (a.s.) when they happen with probability 1; in other words, the set of possible exceptions may be non-empty, but it has probability 0 \citep{Chung}.}
%%%%%%%%%%%%%%%%%%%%%%%%%%%%%%%%%%%%%%%%%%%%%%

%%%%%%%%%%%%%%%%%%%%%%%%%%%%%%%%%%%%%%%%%%
\section{Sequential games with complete information}\label{sec:APS}

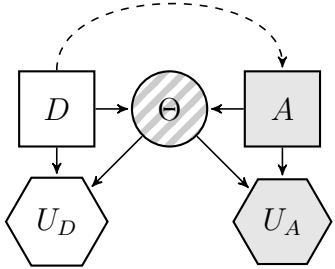
\begin{figure}[htbp]
\centering
\begin{tikzpicture}[->,>=stealth',shorten >=1pt,auto,node distance=1.5cm,
                    semithick]
  \tikzstyle{uncertain}=[circle,pattern=stripes,
                                    pattern color=gray!40,
                                    thick,
                                    minimum size=1.0cm,
                                    draw=black]
  \tikzstyle{utility}=[regular polygon,regular polygon sides=6,
                                    thick,
                                    minimum size=1.0cm,
                                    draw=black,
                                    fill=gray!20]
  \tikzstyle{defensor_utility}=[regular polygon,regular polygon sides=6,
                                    thick,
                                    minimum size=1.0cm,
                                    draw=black,
                                    fill=white]
  \tikzstyle{decision}=[rectangle,
                                    thick,
                                    minimum size=1cm,
                                    draw=black,
                                    fill=gray!20]
  \tikzstyle{defensor_decision}=[rectangle,
                                    thick,
                                    minimum size=1cm,
                                    draw=black,
                                    fill=white]
  \tikzstyle{texto}=[label]
  \node[uncertain](D)   {$\Theta $};
  \node[defensor_decision] (A) [left of=D]  {$D$};
  \node[decision] (B) [right of=D] {$A$};
  \node[defensor_utility]  (C) [below of=A] {$U_D$};
  \node[utility]  (E) [below of=B] {$U_A$};
  \path (A) edge    node {} (D)
            edge    node {} (C)
        (B) edge    node {} (D)
            edge    node {} (E)
        (D) edge    node {} (C)
            edge    node {} (E)
        (A) edge[out=90, in=90, dashed] node {} (B);
\end{tikzpicture}
 \caption{Basic two player sequential defend-attack game.
 White nodes affect solely the Defender; grey nodes affect only the Attacker; 
 striped nodes affect both agents.} \label{fig:tpsdg}
\end{figure}

This section focuses on computational methods for finding equilibria in sequential \textcolor{black}{defend-attack}
%two-stage non-cooperative 
games with complete information. 
 %As an example, consider 
%a company that must determine %decide %what 
%its cybersecurity controls 
 %given that a hacker 
%with the capacity to observe them
%could observe them and launch a distributed denial-of-service (DDoS) attack.
%attack after observing the defensive action of a company deploying cyber %security controls. %In particular, the focus is on simulation based solution methods that could be beneficial for such games with high dimensional decision alternatives and/or uncertain outcomes.
%company deploys cyber security controls and then, having observed the defensive action of a company deploying cyber security controls, a hacker decides whether to launch a distributed denial-of-service (DDoS) attack. %against such company. 
%Consider a sequential game with two agents.
%: the first one makes her decision; then, after having observed such decision, %the second one implements his alternative.
%For example, a company deploys cyber security controls and then, having observed the defensive action, a hacker decides whether to launch a distributed denial-of-service (DDoS) attack. %against such company. 
\renewcommand{\thefootnote}{\arabic{footnote}}
%{\color{black} 
Assume, thus, a Defender ($D$, she)%who 
chooses her defense $d \in \Dcal$, 
\textcolor{black}{ where $\Dcal$ is her
set of feasible alternatives}. After %having observed it
observing this, an Attacker ($A$, he) chooses his attack $a \in \Acal$, \textcolor{black}{ with $\Acal$ his set of 
available alternatives}.  
\textcolor{black}{ The consequences of the interaction for both agents depend on a random 
outcome $\theta \in \Theta$, with $\Theta$ the space of outcomes.}
%Both $\Dcal$ and $\Acal$ are assumed finite, unless noted.
Figure \ref{fig:tpsdg} displays the corresponding bi-agent influence diagram
\citep{Banks:2015} which serves as \textcolor{black}{a} template for later discussions. Arc $D$-$A$ reflects that the Attacker observes the Defender's decision. 
 The agents have their own %{\color{black}
assessment of
%} 
the outcome probability,
respectively $p_D(\theta \given d,a)$ and $p_A(\theta  \given d,a)$. %,
%dependent on $d$ and $a$. 
The Defender's utility $u_D(d, \theta)$ 
is a function of her chosen defense and the 
outcome.
Similarly, the Attacker's utility function is $u_A(a, \theta)$.
%At various places we shall find convenient to redesignate $\theta$ as $\theta$ or $\theta$, as the state is perceived differently by the Attacker or the Defender.

In the standard complete information setup,
the basic game-theoretic solution does not
require $A$ to know $D$'s probabilities and utilities, as he observes
her actions. However, the Defender must know $(u_A,p_A)$, the common knowledge assumption in this case. %complete information. %in this case. 
Then, %\textcolor{black}{for space of outcomes} $\Theta$,
both agents' expected utilities \textcolor{black}{are computed as} %at
 %These methods include computing the expected utilities of both agents', %at node $\Theta$: %in Figure~\ref{fig:tpsdg},
%
$
\psi_A (d,a) = \int u_A (a, \theta )\, p_A(\theta \given d,a)\, \dd \theta $   and
$\psi_D (d,a) = \int u_D (d, \theta )\, p_D(\theta \given d,a) \dd \theta $.
Next, the Attacker's best response to $D$'s 
action $d$, is $a^\opt(d) = \argmax_{a \in \Acal}\, \psi_A(d,a)$.
The optimal attacker response to defense $d$ is then used to find the Defender's optimal action 
%
%\begin{equation*}\label{d_opt}
$d^\opt_\text{GT} = \argmax_{d \in \Dcal}\, \psi_D(d, a^\opt(d))$.
%\end{equation*}
%
In case of multiple best responses for the attacker given some defender action $d$, $a^*(d)$ becomes a set. %} %for some decisions $d$,} 
We can then either use all global optimal alternatives or break the ties.
%Since the assumption of uniqueness of the best response is relatively frequent in the security literature through the strong and weak Stackelberg equilibrium concepts, we break the ties.
In particular, ties are generally \textcolor{black}{addressed by} either choosing the most favorable attack for the defender, leading to a \textit{strong Stackelberg equilibrium (SSE)} (solving $\max _{d\in {\cal D}, a\in a^*(d)}\psi_D (a,d)$); or choosing the worst attack for the defender, leading to a \textit{weak Stackelberg equilibrium} (solving $\max _{d\in {\cal D}} 
\min _{ a\in a^*(d)}\psi_D (a,d)$), see \cite{leitmann}.
%{\color{red}However, we should note that 
%{\color{red}The proposed algorithms can accommodate any tie breaking rule, which results in straightforward handling of multiple best responses.} 
%is straightforward.}
While, in principle, the algorithms 
%to be proposed 
can accommodate any tie breaking rule, in what follows, when necessary, we shall assume that ties are broken in favor of the defender, the standard solution in security games \citep{korzhyk}. \textcolor{black}{ Hence, in case of ties, the element of $a^*(d)$ maximizing the Defender's expected utility will be 
%\textcolor{red}{referred here to as the {\em element of interest} and }
recorded as $a^*(d)$.}
% {\color{red} Thus, in the Algorithms, the element of $a^*(d)$ that maximizes the Defender's expected utility is referred to as the element of interest.} 
The pair $\left( d^\opt_\text{GT},\, a^\opt(d^\opt_\text{GT}) \right)$ is a Nash equilibrium (and, indeed, a sub-game perfect equilibrium \citep{Hargreaves:2004}).
We shall designate $d^\opt_\text{GT}$ as the
game-theoretic \textcolor{black}{ defense under complete information}.
%Note that we use backwards induction for both agents, switching from the Attacker  to the Defender problem as required. 

These sequential games require solving a bilevel optimization problem \citep{bard1991some} which rarely have analytical solutions,
other than %{\color{black}
by explicit enumeration
%} 
in basic models. Even extremely simple instances of bilevel problems have been shown to be NP-hard \citep{jeroslow1985polynomial}; 
\textcolor{black}{
\cite{korzhyk2} provide additional computational complexity results 
in such problems.} Thus, numerical techniques are required. A variety of methods are available, see e.g.\ \cite{Nisan:2007}. In particular, several classical and evolutionary approaches have been proposed, reviewed by \cite{sinha2018review}. When the inner problem adheres to certain regularity conditions, it is possible to reduce the bilevel problem to a single level one, replacing the inner problem by its Karush-Kuhn-Tucker conditions \citep{gordon2012karush}. Then, evolutionary techniques 
may be used to solve this single-level problem, enabling to relax the upper level requirements. 
As this single-level reduction is not generally feasible, other approaches such as nested evolutionary algorithms or metamodeling methods have been proposed. However, most of these approaches lack scalability.
%\textcolor{green}{ an increase in the number of variables in the outer optimization results in an exponential increase of the number of required tasks.}
Yet problems in emerging areas in secure 
artificial intelligence (AI) such as AML \citep{AMLARA}
may require dealing with high dimensional continuous decision spaces, and, consequently, can hardly be solved with standard methods.
%}. 
Some scalable gradient based solution approaches have been recently introduced  \citep{naveiro2019gradient}; however, they are restricted to \textcolor{black}{games
under certainty} or in which expected utilities can be computed analytically. \textcolor{black}{On the other hand, MC simulation methods, see  e.g.\  \cite{Ponsen} and \cite{johanson2012efficient}, are broadly applicable}, as described next.
%we briefly describe next.

%A variety of analytical and numerical  methods are available for solving such games, see e.g.\ \cite{Nisan:2007} or \cite{naveiro2019gradient}.

%%%%%%%%%%%%%%%%%%%%%%%%%%%%%%%%%%%
\subsection{Monte Carlo simulation for games} \label{sec:GT-MC}
Simulation based methods for sequential games typically 
approximate expected utilities using MC %sums 
and, then, optimize with respect to decision alternatives, first to approximate $A$'s best responses,
then to approximate the optimal defense. Algorithm \ref{alg:mcmc} reflects a generic MC approach to solve \textcolor{black}{sequential defend-attack} games %{\color{black} 
when the decision sets $\Acal$ and $\Dcal$
are discrete,\footnote{\textcolor{black}{ In continuous cases, 
%
%When the 
%decision sets 
%, , 
%are continuous, %discretization of the decision sets needs to be introduced
they need to be discretized 
to %appropriately explore 
solve 
the problem to the desired precision, see   
Section \ref{sec:computational_comp}.}}
where 
%} 
$Q$ and $P$ are the sample sizes required to, respectively, approximate the expected utilities $\psi_A (d,a)$ and $\psi_D (d,a)$ to the desired precision. %, as discussed in SM, Section 1.1.
{\footnotesize
\begin{algorithm}[htpb]
\linespread{0.7}\selectfont
%{\footnotesize
\Input{ $Q$, $P$}
\For{ all $d \in \Dcal$   }       
{ 	\For{ all $a \in \Acal$      } 
    {
    	Generate samples $\theta_1, \dots, \theta_Q \sim p_A(\theta \given d,a)$\;
        Compute $\widehat{\psi}_A(d, a) = \frac{1}{Q} \sum_i u_A(a, \theta_i)$\;
   }
   %\textcolor{black}{
   {\color{black} Approximate $a^\opt(d)$ \textcolor{black}{via} %the element of interest in 
   $\argmax_a \widehat{\psi}_A(d,a)$.}
   %}
   \;
   Generate samples $\theta_1, \dots, \theta_P \sim p_D(\theta \given d, a^\opt(d))$\;
   Compute $ \widehat{\psi}_D(d) = \frac{1}{P} \sum_i u_D(d, \theta_i)$\;
}
Compute $\widehat{d}^\opt_\text{GT} = \argmax_d \widehat{\psi}_D(d)$\;
%}
\caption{MC approach for \textcolor{black}{ sequential defend-attack} games with complete information}\label{alg:mcmc}
\end{algorithm}
}

\noindent Convergence of Algorithm \ref{alg:mcmc}, 
sketched in SM Section 1.1, follows under mild conditions and is based on two
applications %(for the inner and outer loops %within the algorithm) 
of a uniform version of the 
strong law of large numbers (SLLN) \citep{Jennrich}. %\textcolor{black}{The optimal attack can be written more generally as the element of $\argmax_a \widehat{\psi}_A(d,a)$ that maximizes $D$'s expected utility. 
%, entailing  
%uniform convergence to the expected utilities.
%as well as to the 
%Attacker's best responses and defender's optimal decision.

%%% PLACE THIS
%However, emerging problems as in adversarial machine learning (Ref) that may have the algorithms %as decision makers result in high dimensional and/or continuous decision spaces cannot be solved %with analytical methods. Cyber security games also have many decision alternatives for the agents %and outcomes. In such cases with high dimensional decision alternative and/or random outcome space %and where computation/estimation of expected utilities is challenging; simulation based methods %could be used, \textcolor{red}{eq Ref1 and Ref2}. MC is among the most widely simulation methods %as its implementation is straightforward and it has well-studied convergence properties.

%The optimal decisions for the expected utility estimates converge to the optimal decisions almost %surely as the sample size diverges to infinity (\cite{shao1989monte}): 
%\begin{center}
%$\widehat{a^{*}_Q(d)} \rightarrow a^{*}(d)$  as $Q\rightarrow \infty$;  
%$\widehat{d^{*}_P} \rightarrow d^{*}$  as $P\rightarrow \infty$.
%\end{center}

%\textcolor{red}{Decide if the following is necessary or pollutes the message-and its link to the %case of mixed strategies::: When $d^{*}$ is not unique, $\widehat{d^{*}_P}$ may not converge to %the global optima due to existence of many limit points. However, $\widehat{d^{*}_P}$ can be used %in practice as a good decision as its expected utility still converges to the optimal expected %utility. (\cite{shao1989monte})}

From a computational perspective, in addition to the cost of the final optimization and $|\Dcal|$ inner loop optimizations, Algorithm 
\ref{alg:mcmc} requires generating $|\Dcal| \times ( |\Acal| \times Q +  P ) $ samples, where $| \boldsymbol{\cdot} |$ designates the cardinality of the corresponding set. 
%or a regression metamodel could be used \cite{metamods}.
\textcolor{black}{MC approaches could be computationally expensive when dealing with decision dependent uncertainties}:  \textcolor{black}{sampling from $p_D(\theta \given d,a)$ and  $p_A(\theta \given d,a)$ is required for each pair ($d$, $a$), %for the Defender's and the Attacker's problem 
entailing loops over the sets} $\Dcal$ and $\Acal$. When 
these are high dimensional, 
%considering the whole decision space as in 4
MC will typically 
be inefficient. APS  %to approximate %game theoretic
%solutions for sequential defend-attack 
mitigates this issue.

%%%%%%%%%%%%%%%%%%%%%%%%%%%%%%%%%%%%%%
%%%%%%%%%%%%%%%%%%%%%%%%%%%%%%%%%%%%%%

%%%%%%%%%%%%%%%%%%%%%%%%%%%%%%%%%%%%%%
%%%%%%%%%%%%%%%%%%%%%%%%%%%%%%%%%%%%%%
%APS requires a positive utility function to get a proper probability density function; adding a large enough number to the utility will do that without changing the nature of the aforementioned distributions. 

%Note that MC simulation approaches Equation \eqref{exp_util} by first estimating the expected utility through an MC average, $\widehat{\psi (x)}$,
%computed using $N$ independent MC samples from $p(\theta \given x)$,
%
%$$\widehat{\psi (x) } =\frac{1}{N} \sum_{i=1}^N u(x,\theta^{(i)}).$$
%
%We then optimize $\widehat{\psi (x) }$ over $x$, \cite{Shao:1988}.

%%%%%%%%%%%%%%%%%%%%%%%%%%%%%%%%%%
\subsection{Augmented probability simulation for games} \label{sec:GT-APS}

%APS approaches to approximate game theoretic solutions for sequential defend-attack games mitigate the issues with high dimensionality

APS solves for expected utility maximization by converting 
the tasks of  estimation and optimization into simulation from an
augmented distribution over the joint space of decisions
and outcomes. %, not requiring a separate optimization step. 
It can be advantageous in problems with expected utility surfaces that are expensive to estimate rendering the optimization step inefficient. 
\cite{BMI:1999} introduced it to solve %maximum expected utility based 
decision analysis problems. 
  \cite{Ekinetal14}
 %\cite{Ekinetal14}, 
%utilized it to solve %decision models with constraints, specifically
%two stage stochastic programs with recourse, whereas 
extend it to solve constrained stochastic optimization models 
%\textcolor{black}{
with recourse, where the second stage problem can be solved analytically and uncertainty is exogenous.
%}
%extended it to constrained domains to deal with stochastic programming problems. 
%\textcolor{black}{
\cite{ekin2017augmented} propose APS models to solve  similar stochastic programs with endogenous uncertainty, whereas \cite{ekin2020augmented} extend it to solve discrete programs. %Similar one stage APS models are used to solve one stage stochastic decision problems in call center staffing \citep{aktekin2016stochastic}, \citep{ekin2021decision} and manufacturing \citep{Ekin:2018}. 
\textcolor{black}{ All these algorithms are designed for problems with a single decision maker and, moreover, with uncertainty only at the first stage: APS is actually used to solve single stage problems (possibly from bilevel problems in which the second stage problem is solved analytically).
Therefore, those APS algorithms cannot be used to solve sequential decision models, in particular sequential games, with two opposing decision makers and endogenous uncertainty
  at both stages, which makes the second stage decision model stochastic.}  %, as in \cite{Ekinetal14}). 
 To the best of our knowledge, this 
  is the first application of APS to solve general two-stage stochastic decision problems (in particular, games).

For a given defender action $d$, consider the augmented
distribution $\pi_A (a, \theta \given d)\propto$ $u_A (a, \theta)\, p_A (\theta \given d, a) $ over $(a,\, \theta)$, thus defined as proportional to the product of $A$'s utility function and his original distribution.
%
%$%$\begin{equation}
%\pi_A (a, \theta \given d) . 
If $u_A (a, \theta)$ is positive and $u_A (a, \theta)\, p_A (\theta \given d, a)$ is integrable, the augmented distribution is well-defined.  
%= \int \pi_A (a, \theta \given d)  d\theta 
  Moreover, %simulating from it solves simultaneously 
%for the expectation of the objective function and its optimization, 
%since 
its  marginal over actions $a$, given by 
$\pi_A(a \given d)= \int \pi_A(a, \theta  \given d) d\theta $,
is proportional to $A$'s expected utility $\psi_A(d, a)$. %= \int 
%u_A (a, \theta)\, p_A (\theta \given d, a) \dd \theta$.
Consequently, 
given $d$, 
the Attacker's
best response can be computed as $a^\opt(d) = \mode\, [\pi_A (a \given d)]$.
%\textcolor{black}{%Note that there could be multiple modes, among which we would need to choose one of the global modes, for which several criteria were mentioned above.}
%\textcolor{black}{As mentioned, 
\textcolor{black}{In case of multiple global modes, we choose the mode of interest by breaking ties via SSE, and record it as $a^\opt(d)$.}
%} 
\textcolor{black}{To avoid pathological cases, we assume that the set of global modes
is finite.}

\textcolor{black}{ Now, assuming that $u_D(d, \theta)$ is positive and $u_D(d, \theta)p_D(\theta \given d,a)$ is integrable, we solve $D$'s problem
by backward 
induction, sampling from the augmented distribution}
%
%\begin{equation}\label{merde1}
$\pi_D(d, \theta \given a^\opt(d)) \propto u_D(d , \theta) p_D(\theta \given d, a^\opt(d))$. 
%
%where we now stress that $\theta$ is perceived by $D$ as 
%$\theta$.
Its marginal $ \pi_D(d \given a^\opt (d))$
in $d$ is proportional to $D$'s expected utility $ \psi_D(d, a^\opt(d))$ and, consequently,
%
%\begin{equation}\label{opt:A}
the game-theoretic solution satisfies
$d_\text{GT}^\opt  = \mode\, \left[\pi_D(d \given a^\opt(d) )\right]$.
\textcolor{black}{ Operationally, this %naturally 
 suggests a two-stage 
\textcolor{black}{nested} approach to sequential %non-cooperative 
games based, at each stage, on 
% on
%\textcolor{red}{can be summarized with the steps of } 
%\textcolor{black}{with each stage including the steps of: 
 i) sampling from the augmented distribution; 
ii)  marginalizing to the corresponding decision variables; and,
iii) estimating the mode of 
the marginal samples.}
%(with a consistent %mode 
%estimator).

%and \cite{romano1988weak}.
%for consistent mode estimation.
%Metropolis algorithms (\cite{metropolis1953equation}) and Gibbs sampling (\cite{roberts1994simple}) are among the widely used MCMC methods, see \cite{gamerman2006markov} for a comprehensive discussion. 
%\subsubsection{Sampling from the augmented distributions} \noindent 

Concerning step (i), 
Markov chain Monte Carlo (MCMC) methods
 \citep{French:2000}   
serve for sampling from the non-standard augmented 
distributions.
These methods construct a Markov chain in the space of the target
distribution (the augmented distributions)  converging to 
the target under mild conditions.
After convergence is detected, 
samples from the chain can be used as approximate target samples.
%from the target. 
Of the various approaches available to construct the 
chains,  
%\textcolor{green}{\ref{ec:gibbs_adg} discusses Gibbs based algorithms; 
%is presented with a discussion of its computational complexity in the . %, f{GibbsGT}.} 
%}
 we adopt versatile Metropolis-Hastings (MH)
variants \citep{chib1995understanding} in 
%due to their applicability in cases of analytically available full conditional distributions that are difficult to sample from. 
%HERE
%Metropolis-Hastings algorithms include an acceptance/rejection step which requires a proposal %(candidate generating) distribution such that the resulting Markov chain is irreducible and %aperiodic.
 Algorithm \ref{alg:MHdefenderAPS2}. This facilitates sampling approximately from $\pi_D(d, \theta \given a^\opt(d))$ (outer APS) to solve $D$'s problem. Within that, $A$'s best response $a^\opt(d)$ is estimated for any $d$ using another APS (inner APS) based
on $\pi_A(a, \theta \given d)$. \textcolor{black}{ Details of  
our  
key MH acceptance/rejection step follow: 
let $d$ and $\theta$ be the current samples in the MH scheme of the outer APS; a candidate $\tilde{d}$ for  
$D$'s decision 
is sampled from a proposal
generating distribution $g_D( \tilde{d} \given d)$;  
%We choose this to be symmetric in the sense that it satisfies $g_D( \tilde{d} \given d) = g_D( d \given \tilde{d})$. { \color{black} However this requirement could be easily removed, including further corrections in the acceptance probability}. 
then, $A$'s problem is \textcolor{black}{solved using} an inner APS to estimate $a^{\opt}(\tilde{d})$; the state $\theta$ is next sampled %given ($\tilde{d}$, $a^{\opt}(\tilde{d})$) 
using $p_D(\theta \given \tilde{d}, a^\opt(\tilde{d}) )$;   
finally, the candidate samples are accepted with probability $\frac{u_D(\tilde{d}, \tilde{\theta}) \cdot g_D( d \given \tilde{d} ) }{u_D(d, \theta ) \cdot g_D( \tilde{d} \given d)}$.}
{\footnotesize 
\begin{algorithm}[h]
\linespread{0.7}\selectfont
\SetKwFunction{Fatk}{solve\_attacker}
\SetKwProg{Fn}{function}{:}{}  

\Fn{\Fatk{$M$, $K$, $d$, $g_A$}}{
\Initialize{$a^{(0)}$ }
Draw $\theta^{(0)} \sim p_A(\theta \given d, a^{(0)})$\;
\For(\Comment*[f]{Inner APS}){$i=1$ \KwTo $M$ } 
{
    Propose new attack $\tilde{a} \sim g_A(\tilde{a} \given a^{(i-1)})$.\;
    Draw $\tilde{\theta} \sim p_A(\theta \given d, \tilde{a})$\;
    Evaluate acceptance probability
    $
    \alpha = \min \left \lbrace 1, \frac{u_A\left( \tilde{a}, \tilde{\theta} \right) \cdot g_A \left(a^{(i-1)} \vert \tilde{a}\right) }{u_A\left(a^{(i-1)}, \theta^{(i-1)} \right) \cdot g_A\left(\tilde{a} \given a^{(i-1)}\right )} \right \rbrace
    $\;
    With probability $\alpha$ set $(a^{(i)},
    \theta^{(i)})=(\tilde{a},  \tilde{\theta})$.
        Otherwise, set $(a^{(i)}, \theta^{(i)})
        = ( a^{(i-1)}, \theta^{(i-1)})$.\;
    %Evaluate practical MCMC convergence\;
    }
Discard first $K$ samples and \textcolor{black}{compute $a^\opt(d)$ based on the mode(s) of $\{a^{(K+1)},...,a^{(M)}\}$}\;
%{\color{black} element of interest (mode) from }
%\textcolor{red}{mode of rest of draws $\{a^{(K+1)},...,a^{(M)}\}$} as $a^\opt(d)$.\;
\textbf{return} $a^\opt(d)$\;
}
\Input{$M$, $K$, $N$, $R$, $g_D$ and $g_A$ proposal distributions } 
\Initialize{$d^{(0)}$, $a^\opt(d^{(0)})$ = \Fatk{$M$, $K$, $d^{(0)}$, $g_A$} }
Draw $\theta^{(0)} \sim p_D(\theta \given d^{(0)}, a^\opt(d^{(0)}) )$\;
\For(\Comment*[f]{Outer APS}){$i=1$ \KwTo $N$} {
    Propose new defense $\tilde{d} \sim g_D(\tilde{d} \given d^{(i-1)})$\;
    $a^\opt(\tilde{d}) =$ \Fatk{$M$, $K$, $\tilde{d}$, $g_A$} if not previously computed\; 
    Draw $\tilde{\theta} \sim p_D(\theta \given \tilde{d}, a^\opt(\tilde{d}) )$.\;
    Evaluate acceptance probability
    $
    \alpha = \min \left \lbrace 1, \frac{u_D\left(\tilde{d}, \tilde{\theta}\right) \cdot g_D\left( d^{(i-1)} \vert \tilde{d} \right) }{u_D \left(d^{(i-1)}, \theta^{(i-1)} \right) \cdot g_D \left(\tilde{d} \given d^{(i-1)} \right)} \right \rbrace
    $\; 
    With probability $\alpha$ set $(d^{(i)},
    a^\opt(d^{(i)}), \theta^{(i)})= 
   ( \tilde{d}, a^\opt(\tilde{d}), \tilde{\theta})$.
    Otherwise, set $(d^{(i)}, \textcolor{black}{a^\opt(d^{(i)}),} \theta^{(i)})=
    ( d^{(i-1)}, \textcolor{black}{a^\opt(d^{(i-1)}),} \theta^{(i-1)})$.\;
    }
Discard first $R$ samples and estimate mode of
$\{d^{(R+1)},...,d^{(N)}\}$. Record it as $\widehat{d}^\opt _\text{GT}$.\;
\caption{MH based APS for \textcolor{black}{ sequential defend-attack} games. Complete information.} \label{alg:MHdefenderAPS2}
\end{algorithm}
}
%%%%%%%%%%%%%%%%%%%%%%%%%%%
 
 \noindent Algorithm \ref{alg:MHdefenderAPS2}  thus defines a
 Markov chain %in $(d, \theta)$ such that
 $(d^{(n)},\theta^{(n)}) \overset {d}{\longrightarrow} $ %(d^{(\infty)},\theta^{(\infty)},a^{(\infty)})$ as $N\rightarrow\infty$, with $(d^{(\infty)},\theta^{(\infty)},a^{(\infty)})$ having distribution
 $\pi_D(d, \theta \given a^*(d))$.\footnote{\textcolor{black}{The symbol
 $\overset {d}{\longrightarrow}$ represents convergence in distribution \citep{Chung}}.} 
   \textcolor{black}{ Step ii) in the \textcolor{black}{APS} scheme is trivial. For step iii)}, \textcolor{black}{ we use recent work 
  in multivariate multiple mode  estimation, see the review in   \cite{chacon},
  %and the approach in \cite{chen},   
  and then break ties when necessary}.
%the classical unimodal estimation results of \cite{romano1988weak} when breaking ties. For a more general multimodal multidimensional estimation, }.
%  mode estimation as in Romano (1988) which focuses on unimodal distributions
% there is recent work in multimodal  multidimensional estimation excellently reviewed in \cite{chacon} including  the approach of  \cite{chen}}.
 
\textcolor{black}{ 
Proposition \ref{prop:conv_nested} requires  
the following four
%pretty  general 
%reasonable 
conditions for the convergence of
Algorithm \ref{alg:MHdefenderAPS2} output 
to $d^\opt_\text{GT}$,
as proved in \ref{apdx:proofs}:
\begin{enumerate}[a).]
\item \textcolor{black}{The Attacker's and Defender's
utility functions 
are positive and continuous in 
$(a,\theta )$ and $(d, \theta )$, respectively.} %Positivity allows for the definition  of the augmented distributions; it is easily achieved taking into account affine uniqueness properties of utility functions \citep{French:2000}. Continuity is also natural and leads to continuous expected utility functions  enabling the existence of optima, whenever 
\item \textcolor{black}{The Attacker's and Defender's decision sets, respectively 
$\mathcal{A}$ and $\mathcal{D}$, are compact.} %,  a usual condition in optimization settings to ensure the existence of optima; 
\item \textcolor{black}{The Attacker's and Defender's probability distributions of the outcomes are continuous in $a$ and $d$ respectively, and \textcolor{black}{are positive in $(d,a)$.}
%verify $p_A(\theta |\, d,a) > 0$ and $p_D(\theta |\, d, \, a) > 0$.
} %These are also quite general and guarantee continuity of the best responses. 
\item \textcolor{black}{The proposal generating distributions, $g_A$ and $g_D$,
have support
$\mathcal{A}$ and $\mathcal{D}$, respectively.} %These are standard sufficient conditions to ensure convergence of MH samplers \citep{smith1993bayesian}. 
%Specific choices are described below.   
\end{enumerate}
}
\vspace{-1cm}
%\textcolor{black}{  %Then,
%; in other words, the set of possible exceptions may be non-empty, but 
%has probability 0 .
%The mode of the corresponding marginal augmented distributions is assumed to be estimated consistently as in e.g.\ \cite{romano1988weak}.
 %In particular, to solve the Defender's problem, a Markov chain is defined such that %$(d^{(N)},\theta^{(N)}, a^{(N)}) \overset {D}{\longrightarrow} %(d^{(\infty)},\theta^{(\infty)},a^{(\infty)})$ as $N\rightarrow\infty$, with %$(d^{(\infty)},\theta^{(\infty)},a^{(\infty)})$ having distribution $\pi_D(d, \theta, a)$ where %$\overset {D}{\longrightarrow}$ represents convergence in distribution. %This results in a geometric convergence rate.  %\textcolor{red}{This assumes samples of APS for defenders decision space gives $a^{*}$--should we include mode discussion in the proof or not}
%the convergence of the algorithm to the optimal decision, with the proof in Appendix \ref{apdx:proofs}.
%
\textcolor{black}{
\begin{prop}\label{prop:conv_nested}
Assume that the Attacker's and Defender's utility functions, decision sets, probability distributions and proposal generating distributions satisfy conditions a), b), c) and d).
%are positive and continuous in, respectively,
%$(a,\theta )$ and $(d, \theta )$; 
%$p_A(\theta |\, d,a)$,
%$p_D(\theta |\, d, \, a) > 0$ $\forall a, d, \theta$ 
%and are continuous in $a$ and $d$; 
%u_A(a, \theta)p_A(\theta |\,  d,a)$
%and $u_D(d, \theta) p_D(\theta |\, d, \, a)$ are integrable;
%$\mathcal{A}$ and $\mathcal{D}$  
%{\color{black} 
%are %either discrete
%sets 
%or %discretized continuous variables represented with 
%compact sets in $\mathbb{R}^n$; %or discrete sets
Then, if the best response sets $a^\opt(d)$ are finite for each $d$, Algorithm \ref{alg:MHdefenderAPS2} defines a Markov chain
with stationary distribution $\pi_D(d, \theta \given a^\opt(d))$ and a consistent mode estimator 
based on its marginal samples in $d$ converges a.s.\ to $d^\opt_\text{GT}$.\footnote{ \textcolor{black}{ We use standard probability theory 
terminology \citep{Chung} to
 refer to events happening almost surely (a.s.) when they occur with probability 1.}}
% its marginal distribution on the Defender's decision space, $\pi_D(d)$, coincides with $d^\opt_\text{GT}$. 
\end{prop}
%}
}
\noindent 
\textcolor{black}{The positivity of the utility functions $u_A$ and $u_D$ allows for the definition  of the augmented distributions; it is easily achieved taking into account affine uniqueness properties of utility functions \citep{French:2000}. Continuity of these utility functions leads to continuous expected utility functions enabling the existence of optima, when assumption b) holds, a standard condition in optimization. Condition c) is
also general and allows for the continuity of best responses.
Finally, assumption d) provides standard sufficient conditions
to ensure convergence of MH samplers \citep{smith1993bayesian}. }

\textcolor{black}{Our proposal generating distributions
$g_D$ and $g_A$ in 
Algorithm 2 are $t$  distributions centered at the current solutions \citep{gamerman2006markov}, when dealing with 
 continuous decisions. 
When facing discrete decisions,} these are displayed
in a circular list and generated from neighboring states with equal probability.
Chain convergence (to discard the first $K$ or $R$ samples) 
may be assessed with various statistics like Brooks-Gelman-Rubin's  \citep{brooks1998assessing}.  Convergence of the Markov chain is at a geometric rate as a function of the minimum and maximum utility values,
%This geometric rate of convergence depends on $b$, a function of the bounds of the utility, $u_0$ and $u_1$.
%Practically, these bounds can be chosen with respect to the maximum and minimum values of the %utility function. 
see SM Section 2.1. Once the chain is judged to have converged, initial samples are discarded as burn-in and the remaining simulated values are used as an approximate sample from the distribution of interest. In particular,
the marginal draws $d^{(R+1)},..., d^{(N)}$ would correspond to 
a sample from, approximately, the marginal distribution, $\pi_D(d \given a^*(d))$.  %Using the strong law of large numbers, such almost sure convergence is written as: $\argmax \{\# a; a \in \{a_{(K+1)},...,a_{(M)}  \} \} \rightarrow mode[\pi_A(a \given d)]= a^{*}(d)$  and 
%$\argmax \{\# d; d \in \{d_{(R+1)},...,d_{(N)}  \} \} \rightarrow mode[\pi_D(d \given a^\opt(d))]= d^\opt_\text{GT}$.
\textcolor{black}{ The sample modes must be estimated with 
a consistent multiple mode estimator in the sense of 
\cite{chen}. We then select among them with the tie-breaking 
criteria mentioned above. See  
%While consistent mode estimation is a reasonably well-studied problem, see \cite{romano1988weak} and
 \cite{chacon} and \cite{romano1988weak} for additional information and references
 in  mode estimation.} 

\textcolor{black}{In comparison to Algorithm 1,}
Algorithm \ref{alg:MHdefenderAPS2} removes the loops over both
${\cal D}$ and ${\cal A}$. Thus, its complexity does not depend on the dimensions of those sets, %{\color{black}
providing
%} 
an intrinsic advantage over MC approaches in problems with large or continuous spaces, as
APS can be directly used with continuous decision sets without
the need of discretization.
Indeed, Algorithm \ref{alg:MHdefenderAPS2} requires
$ N \times (2 \times M + 3) + 2M + 2$ samples plus the cost of convergence checks and 
(at most) $N + 1$ mode approximations, \textcolor{black}{ where $M$ and $N$ 
are, respectively, the number of MCMC iterations for the attacker's and defender's APS required to achieve the desired 
precision.} %It could be preferred when facing problems where the cardinality of the decision %spaces are large or the decisions are continuous.
%$ 2 \times (|\Dcal| \times M +  N)$ samples plus the cost of convergence checks and $|\Dcal| + 1$ mode approximations.

%%%%%%%%%%%%%%%%%%%%%%%%%%%%%%%%%%%%%%%%%%%%%%%%%%%%%%%%%
\paragraph{Example 1}
\textcolor{black}{ Consider the \textcolor{black}{following} simple security game.}  
%referring to an organization that needs to determine its security protocol. %The decision alternatives are discrete: %, they could be in the form of a safe but costly option, or through cheaper options with weaker protection levels, rendering business performance increasingly at risk.
%After structuring the problem, we consider two different assumptions regarding the Defender beliefs over the Attacker beliefs and preferences. First, we discuss the basic game theoretic approach under common knowledge, in which the Defender knows the Attacker utility function $u_A$ and probability evaluations $p_A$. We perform a sensitivity analysis to evaluate the impact of this assumption over the Defender's decision and utility function. Then, we illustrate the ARA which relaxes the common knowledge assumption via evaluation of the Attacker’s uncertainty through $U_A$ and $P_A$ from the Defender's perspective and the solution of the resulting decision problem.
 %%%%%%%%%%%%%
% \subsection{Basic elements}
%We first structure the problem.  %Problem Definition
An organization ($D$) has to choose among ten
security protocols: $d=0$ (no extra defensive action); 
$d=i$  (level $i$ protocol
with increasing protection), $i=1,\dots,8$;
$d=9$ (safe but cumbersome protocol). $A$ has two alternatives: attack ($a=1$) or 
not ($a=0$). Successful (unsuccessful) attacks are denoted $\theta = 1$ ($\theta = 0$). 
When there is no attack, $\theta = 0$.

{\footnotesize
\begin{table}[htbp]
\centering
\begin{subtable}[b]{0.2\textwidth}
\centering
\setlength\extrarowheight{-3pt}
\begin{tabular}{lrr}
\toprule
& \multicolumn{2}{c}{$\theta$} \\
\cmidrule(l){2-3}
$d$ &     0 &     1 \\
\midrule
0 &  0.05 &  7.05 \\
1 &  0.10 &  7.10 \\
2 &  0.15 &  7.15 \\
3 &  0.20 &  7.20 \\
4 &  0.25 &  7.25 \\
5 &  0.30 &  7.30 \\
6 &  0.35 &  7.35 \\
7 &  0.40 &  7.40 \\
8 &  0.45 &  7.45 \\
9 &  0.50 &  7.50 \\
\bottomrule
\end{tabular}
\caption{}\label{tab:cd}
\end{subtable}
\begin{subtable}[b]{0.2\textwidth}
\centering
\setlength\extrarowheight{-3pt}
\begin{tabular}{lrr}
\toprule
& \multicolumn{2}{c}{$a$} \\
\cmidrule(l){2-3}
$d$ &     0 &     1 \\
\midrule
0 &  0.0 &  0.50 \\
1 &  0.0 &  0.40 \\
2 &  0.0 &  0.35 \\
3 &  0.0 &  0.30 \\
4 &  0.0 &  0.25 \\
5 &  0.0 &  0.20 \\
6 &  0.0 &  0.15 \\
7 &  0.0 &  0.10 \\
8 &  0.0 &  0.05 \\
9 &  0.0 &  0.01 \\
\bottomrule
\end{tabular}
\caption{}\label{tab:prob}
\end{subtable}
\begin{subtable}[b]{0.2\textwidth}
\centering
\setlength\extrarowheight{-3pt}
\begin{tabular}{lrr}
\toprule
& \multicolumn{2}{c}{$\theta$} \\
\cmidrule(l){2-3}
$a$ &     0 &     1 \\
\midrule
0 &  0.00 &  0.00 \\
1 & -0.53 &   1.97 \\
\bottomrule
\end{tabular}
\caption{}\label{tab:ca}
\end{subtable}
\begin{subtable}[b]{0.2\textwidth}
\centering
\setlength\extrarowheight{-3pt}
\begin{tabular}{lrr}
\toprule
$d$ & $\alpha_d$ & $\beta_d$ \\
\midrule
0 &   50.0 &  50.0 \\
1 &   40.0 &  60.0 \\
2 &   35.0 &  65.0 \\
3 &   30.0 &  70.0 \\
4 &   25.0 &  75.0 \\
5 &   20.0 &  80.0 \\
6 &   15.0 &  85.0 \\
7 &   10.0 &  90.0 \\
8 &    5.0 &  95.0 \\
9 &    1.0 &  99.0 \\
\bottomrule
\end{tabular}
\caption{}\label{tab:dist}
\end{subtable}
\caption{\subref{tab:cd} Def. net costs; \subref{tab:prob} Successful attack probs.; \subref{tab:ca} Att. net benefits; \subref{tab:dist} Beta dist. parameters}
\end{table}
}

%%%%%%%%%%%%%%%%
\subparagraph{Defender non strategic judgments.}
Table \ref{tab:cd} presents costs $c_D$ associated with each decision and outcome, based on a $7$M\euro \, business valuation;  and $0.05$M\euro \, base security cost plus $0.05$M\euro
\, per each security level increase.
Upon successful attack, $D$ loses the entire  
business value. 
The probability $p_D (\theta=1 \given d,a)$ of  
successful attack
%given $d$ and $a$ 
is in Table \ref{tab:prob} 
(complementary values for unsuccessful attacks).
% (e.g., $p_D (\theta  = 0 \given a = 1, d = 2)= 1-0.35$).
$D$ is constant risk averse in costs, with utility strategically
equivalent to $u_{D} (c_{D}) = - \exp{ (0.4 \times c_{D}) }$.
%and $c = 0.4$.
%%%%%%%%%%%%%%%%%%%%%%%%
\subparagraph{Attacker judgments.}
%
%Consider the Attacker problem.
The average attack cost is $0.03$M\euro . The average attack benefit
 is $2$M\euro . An unsuccessful attack has an extra cost of $0.5$M \euro . Table \ref{tab:ca} presents the Attacker's net benefit $c_A (a , \theta)$.
$D$ thinks that $A$ 
is constant risk prone over benefits, with 
utility strategically equivalent to $u_{A} (c_{A}) = \exp{ (e \times c_{A}) }$, with
$e>0$. 
%%%%%%%%%%%%%%%%%%%%%%%%%%%%%%%%%%%%%%%%%
\subparagraph{Complete information case.} 
To fix ideas, assume  
$p_A (\theta =1 \given d,a ) = p_D (\theta =1 \given d,a )$ (Table \ref{tab:prob})
and $e=1$. 
\begin{figure}[htbp]
\begin{subfigure}{0.4\textwidth}
\centering
\includegraphics[width=\textwidth]{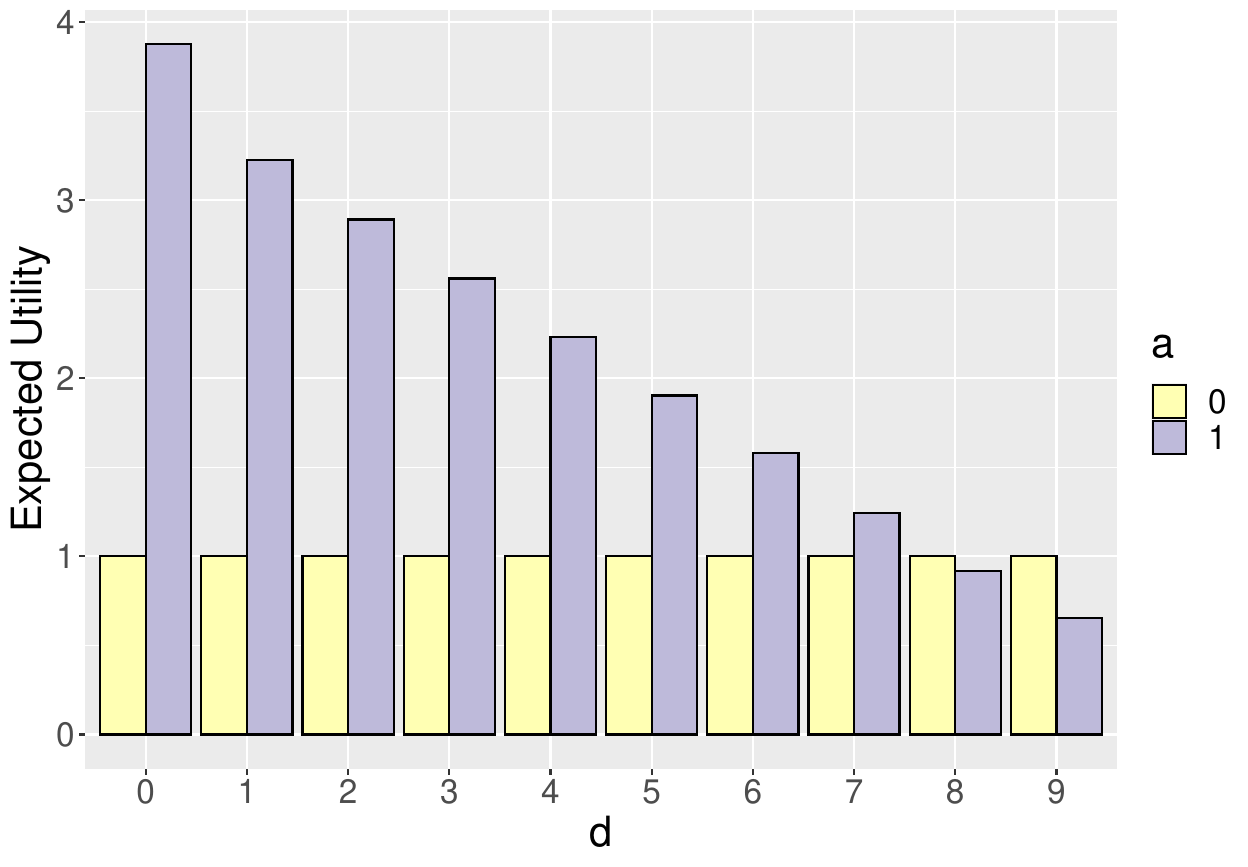}
\caption{MC solutions}\label{fig:MC-GTA}
\end{subfigure}
\hfill
\begin{subfigure}{0.4\textwidth}
\centering
\includegraphics[width=\textwidth]{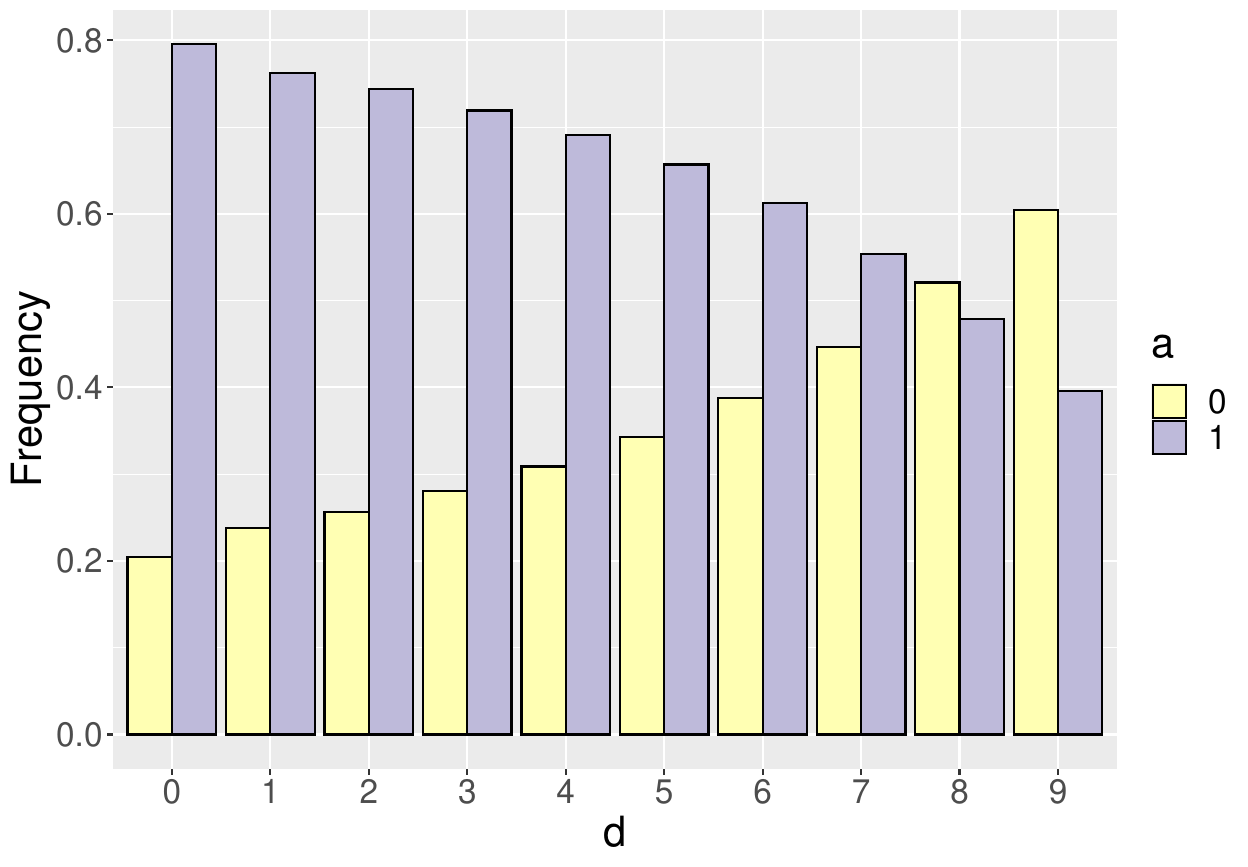}
\caption{APS solutions}\label{fig:APS-GTA}
\end{subfigure}
\caption{Attacker problem solutions for each defense}\label{fig:GTA}
\end{figure}
\noindent MC and APS approximate optimal decisions using Algs.
\ref{alg:mcmc} and \ref{alg:MHdefenderAPS2}, respectively. 
Fig. \ref{fig:MC-GTA} represents MC estimates
of $A$'s expected utility for
each $d$ and $a$. 
For each $d$, the best response 
$a^*(d)$   is the maximum expected utility
alternative;
e.g.\ for $d = 5$, 
$A$'s optimal decision is to attack; for $d = 8$, he should not attack.
 Fig. \ref{fig:APS-GTA} represents the frequencies of marginal samples of %results of sampling from the marginal in 
$a$ from the augmented distribution $\pi_A(a, \theta \given d)$ for each $d$. Its mode coincides with the optimal attack.  
\textcolor{black}{ For $d\leq  7 $,
%(level 7 protocol),
the Attacker should attack.} %, under the common knowledge assumption.
With stronger defenses ($d\geq 8 $), the mode is $a=0$ and attack is not advised. The Attacker's best 
responses $a^*(d)$ %for each defense $d$ are
 thus coincide for both approaches.

\begin{figure}[htbp]
\begin{subfigure}{0.4\textwidth}
\centering
\includegraphics[width=\textwidth]{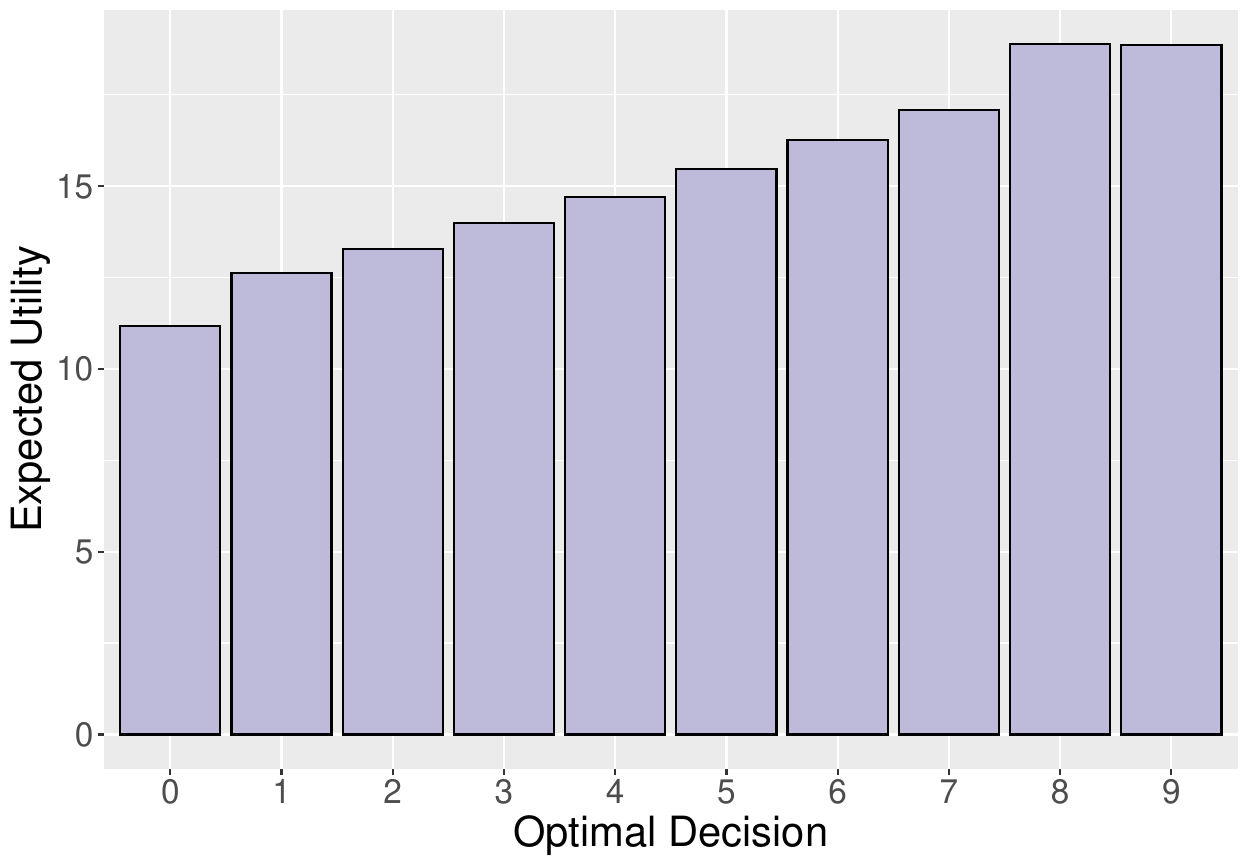}
\caption{MC solution}\label{fig:MC-GTD}
\end{subfigure}
\hfill
\begin{subfigure}{0.4\textwidth}
\centering
\includegraphics[width=\textwidth]{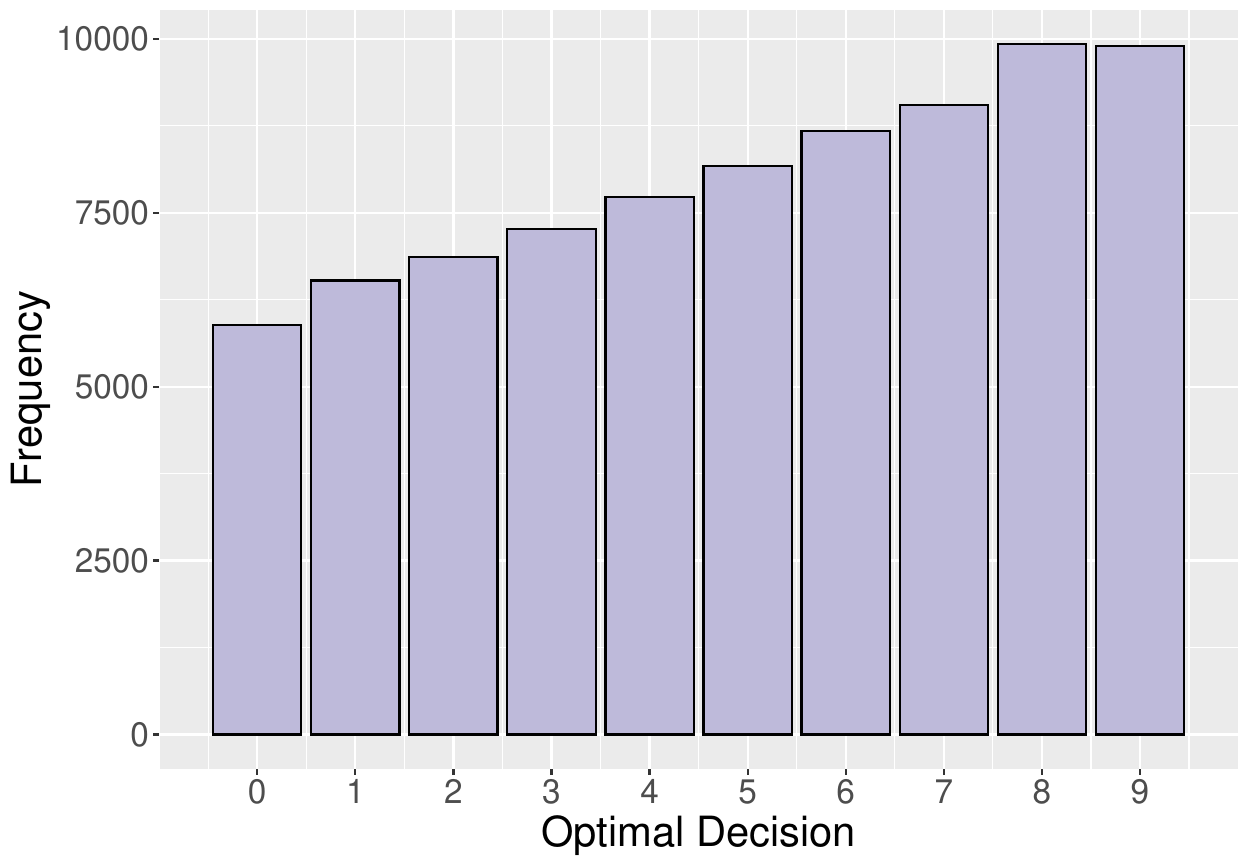}
\caption{APS solution}\label{fig:APS-GTD}
\end{subfigure}
\caption{Solutions of Defender problem}\label{fig:GTD}
\end{figure}

Armed with $a^*(d)$, the optimal defense is computed, again both using MC and APS. Figure \ref{fig:MC-GTD} presents the MC estimation of $\psi_D(d, a^*(d))$ for each $d$. Figure \ref{fig:APS-GTD} shows sample frequencies from the marginal augmented distribution $\pi_D(d \given a^*(d))$. 
%{\color{black} 
\textcolor{black}{ Both methods agree on  
$d^*_\text{GT}=8$, with $d=9$ a close competitor.}
%} 
It could be argued that finding the exact optimal decision is not  that crucial as 
the expected utilities 
%{\color{black} 
of both decisions are close.  Moreover,  
given such closeness,
%},
it is even challenging to find the exact $d^*_{GT}$. 
%that $d=8$ is the mode of the augmented distribution. 
APS is helpful in checking that, indeed, $d=8$ is  optimal.
%{\color{black}
For this, we %just 
sample $d$ from a
distribution more peaked around the optimal decision by just raising the marginal augmented distribution to a 
power. SM Section 2.2 \textcolor{black}{and 4.1} present the details. %as utilized in Section 3.1
%}
\hfill  $\triangle$ 
%it provides the distribution $\pi_D(d \given a^*(d))$ as part of the solution, delivering sensitivity analysis at no extra cost. %See also Section \ref{sec:sens_anal}.
\vspace{0.07in}

\noindent \textcolor{black}{  This emphasizes another advantage of APS: despite flatness of the expected utility, APS finds the optimal solution with a relatively small %little 
additional computational cost while displaying sensitivity.}

%%%%%%%%%%%%%%%%%%%%%%%%%%%%%%%%%%%%%%
\subsection{Sensitivity analysis} \label{sec:sens_anal}
%$D$'s 
The Defender's
judgments, expressed through $(u_D, p_D)$,
could be argued to be %reasonably 
properly  assessed, 
as she is the supported agent in the game.
However,
as argued in \cite{KEENEY},
our knowledge about $(u_A, p_A)$ may not be that precise: it would require
$A$ to reveal his beliefs and preferences. This is doubtful in domains
such as \textcolor{black}{ security and} cybersecurity where information is concealed and hidden to adversaries.

%since the Defender is receiving support for her decision, we would know $(u_D, p_D)$ reasonably well. 
%{\color{black} 
One could conduct a sensitivity analysis to mitigate this, considering that $A$'s preferences and beliefs are modeled through classes of utilities $u \in \Ucal_A$ and probabilities $p \in \Pcal_A$ summarizing the information available to $D$.
%}  
%possibly obtained from leakage, earlier interactions or informants.
The stability of the proposed $d_\text{GT}^*$
could be assessed  by comparing NE defenses $d^\opt_{u,p}$ computed for each pair $(u,p)$ %. %could be computed with the techniques from Sections \ref{sec:GT-MC} or \ref{sec:GT-APS}. 
based on criteria proposed in the areas of sensitivity analysis in decision making under uncertainty 
\citep{David} and robust Bayesian analysis \citep{Rios:2012}.
Of those, we 
use
the regret $r_{u,p}(d^\opt_\text{GT})= \psi_D(d^\opt_\text{GT}, a^\opt(d^\opt_\text{GT})) -  \psi_D(d_{u,p}^\opt, a^\opt(d_{u,p}^\opt))$, 
\textcolor{black}{ for $(u, p) \in \Ucal_A \times \Pcal_A$, }
since it reflects 
the loss in expected utility %if we choose 
due to choosing $d^*_\text{GT}$ %and 
instead of the defense $d^\opt_{u,p}$ associated with the actual judgments. 
%This serves to control such losses.
%the actual judgments are  $(u, p) \in \Ucal_A \times \Pcal_A$,
%for which we should choose 
%$d^\opt_{u,p}$. 
Small values of $r_{u,p}(d^\opt_\text{GT})$ would indicate robustness with respect to the Attacker's utility and probability:
any pair $(u,p)\in \Ucal_A \times \Pcal_A$ could be chosen with no significant changes in the attained expected utilities
and $d_\text{GT}^*$ is thus robust. Otherwise, the 
relevance of the proposed NE defense $d^\opt_\text{GT}$ should be criticized %{\color{black} 
given the %important 
potential losses, 
%}
and further investigated.
%At a deeper level, this also questions the appropriateness of $(u_A, p_A)$,
%serving to criticize the common knowledge assumption.
Operationally, a {\em threshold} 
%{\color{black}
indicating a maximum acceptable loss in expected utility
%} 
should be specified by the decision maker
%to determine if common knowledge assumptions for the attacker parameters hold. This is
 (as Algorithm \ref{alg:robustgt} sketches).
%, in which \textit{threshold} is set based 
%on the maximum acceptable loss tolerated.

%The exploratory nature of sensitivity analysis does not require large sample sizes as in Algorithm \ref{MHdefenderAPS2} thus enabling us to allocate more resources in exploring a larger sample of $(u, p)$'s.
%
{\footnotesize  
\begin{algorithm}[h!]
\linespread{0.7}\selectfont
\Input{$d^\opt_\text{GT}$, $\Ucal_A$, $\Pcal_A$, \textcolor{black}{$V$}, {\em threshold}}
\For{$i=1$ \KwTo $V$} {
    Randomly sample $u$ from $\Ucal_A$ and $p$ from $\Pcal_A$\;
    Compute $d^\opt_{u,p}$ using Algorithm \ref{alg:MHdefenderAPS2}\;
    Compute $r_{u,p}(d^\opt_\text{GT})$\;
    \If{$r_{u,p}(d^\opt_\text{GT}) > $ \text{threshold} }{
        Robustness requirement not satisfied\;
        Stop
    }
}
Robustness requirement satisfied.
\caption{Robustness assessment of solutions for games with complete information}\label{alg:robustgt}
\end{algorithm}
}
%%%%%%%%%%%%%%%%%%%%%%%%%%%%%%%%%%%%%%%%%%%%%%%%
\paragraph{Example 1. (cont.)}
We check the robustness of $d_\text{GT}^*$. % with respect to the utility and probability 
%assumptions. 
The optimal defense is computed for $10,000$ perturbations
of $u_A(c_A)$
(sampling $e' \sim {\cal U}(0,2)$ and using
$u'_{A} (c_{A}) = \exp{ (e'\times c_{A}) )}$
%Regarding $p_A(\theta \given d, a)$, 
and the probability $p_A(\theta \given d, a = 1) $ of successful attack in the event of an attack for each $d$
(sampling from a Beta distribution with mean equal to the 
original value and variance $0.1$\% of the corresponding mean for each $d$).
%Obviously, when there is no attack, $p_A(\theta = 1 \given d, a = 0) = 0$ for all $d$.
\begin{figure}[htbp]
\centering
\includegraphics[width=0.35\textwidth]{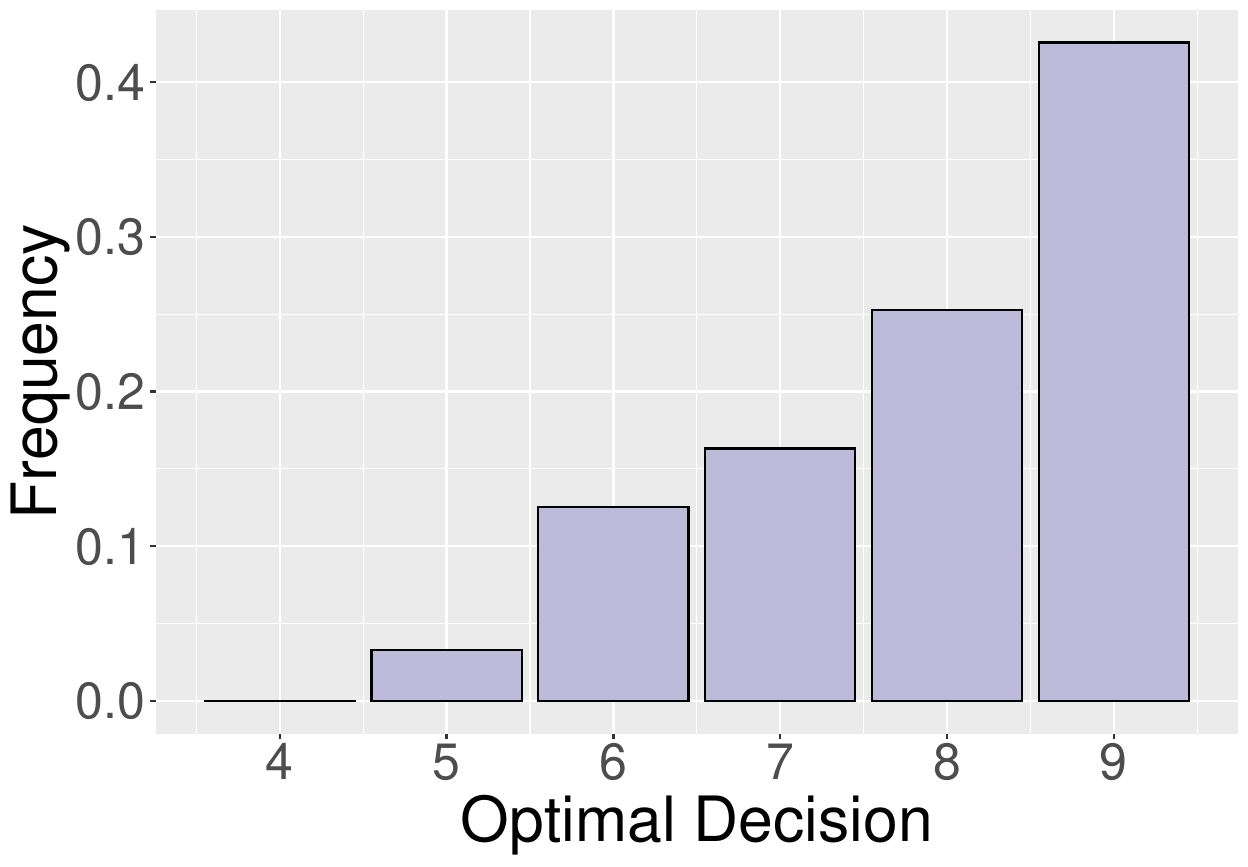}
\caption{Sensitivity analysis of the solution of the game with complete information}\label{fig:SA}
\end{figure}
Figure \ref{fig:SA} reflects the frequency with which each $d$ is %found
%to be 
optimal.
The proposed solution $d^*_\text{GT}=8$ emerges only $25\%$ of the time
as optimal. Moreover, %it is unstable as 
%{\color{black} 
small
perturbations in the utilities and probabilities lead to
 other %potentially
 optimal 
solutions: $d=9$ and $d=7$ respectively emerge $42\%$ 
and $16\%$ of the times as optimal.
%}. are observed $33\% $ of the time, with
More importantly, large variations in optimal expected utilities are observed, with a maximum regret of $42.5 \%$ of the total optimal expected utility. 
To sum up, $d_\text{GT}^* = 8$ 
is clearly sensitive to changes 
in $u_A$  and $p_A$:   
%{\color{black} 
%Indeed, $d_\text{GT}^*$ 
it is not robust for even relatively high regret thresholds for the loss in expected utility.%, such as 30\% of the optimal expected utility.
%, would be overpassed.
%}
\hfill $\triangle$
\vspace{0.03in}

%\noindent This is addressed next by relaxing the complete information assumption. 
%This paper will relax the complete information assumption to address this issue.
%
%%%%%%%%%%%%%%%%%%%%%%%%%%%%%%%%%

%%%%%%%%%%%%%%%%%%%%%%%%%%%%%%%%%
%%%%%%%%%%%%%%%%%%%%%%%%%%%%%%%%%
\section{Sequential games with incomplete information: ARA}\label{sec:ARA}

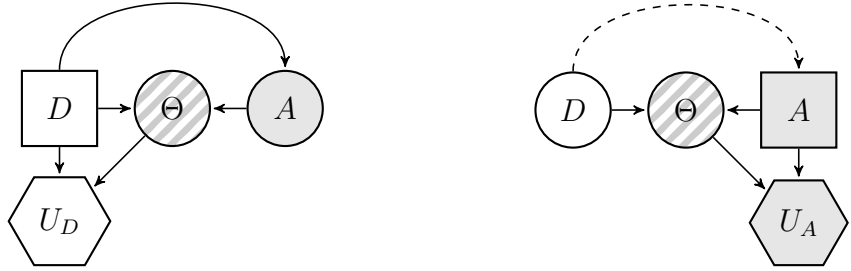
\begin{figure}[htbp]
\begin{subfigure}[t]{0.49\textwidth}
\centering
\begin{tikzpicture}[->,>=stealth',shorten >=1pt,auto,node distance=1.5cm,
                    semithick]

  \tikzstyle{uncertain}=[circle,
                                    thick,
                                    pattern=stripes,
                                    pattern color=gray!40,
                                    minimum size=1.0cm,
                                    draw=black]
  \tikzstyle{attacker_uncertain}=[circle,
                                    thick,
                                    minimum size=1.0cm,
                                    draw=black,
                                    fill=gray!20]
  \tikzstyle{utility}=[regular polygon,regular polygon sides=6,
                                    thick,
                                    minimum size=1.0cm,
                                    draw=black,
                                    fill=gray!20]
  \tikzstyle{defensor_utility}=[regular polygon,regular polygon sides=6,
                                    thick,
                                    minimum size=1.0cm,
                                    draw=black,
                                    fill=white]
  \tikzstyle{decision}=[rectangle,
                                    thick,
                                    minimum size=1cm,
                                    draw=black,
                                    fill=gray!20]
  \tikzstyle{defensor_decision}=[rectangle,
                                    thick,
                                    minimum size=1cm,
                                    draw=black,
                                    fill=white]

  \node[uncertain](D)   {$\Theta $};
  \node[defensor_decision] (A) [left of=D]  {$D$};
  \node[attacker_uncertain] (B) [right of=D] {$A$};
  \node[defensor_utility]  (C) [below of=A] {$U_D$};

  \path (A) edge    node {} (D)
            edge    node {} (C)
        (B) edge    node {} (D)
        (D) edge    node {} (C)
        (A) edge[out=90, in=90]  node {} (B);
\end{tikzpicture}
\caption{Defender's decision problem.}\label{fig:ddp}
\end{subfigure}
\hfill
\begin{subfigure}[t]{0.49\textwidth}
\centering
\begin{tikzpicture}[->,>=stealth',shorten >=1pt,auto,node distance=1.5cm,
                    semithick]

  \tikzstyle{uncertain}=[circle,
                                    thick,
                                    pattern=stripes,
                                    pattern color=gray!40,
                                    minimum size=1.0cm,
                                    draw=black]
  \tikzstyle{defensor_uncertain}=[circle,
                                    thick,
                                    minimum size=1.0cm,
                                    draw=black,
                                    fill=white]
  \tikzstyle{utility}=[regular polygon,regular polygon sides=6,
                                    thick,
                                    minimum size=1.0cm,
                                    draw=black,
                                    fill=gray!20]
  \tikzstyle{defensor_utility}=[regular polygon,regular polygon sides=6,
                                    thick,
                                    minimum size=1.0cm,
                                    draw=black,
                                    fill=white]
  \tikzstyle{decision}=[rectangle,
                                    thick,
                                    minimum size=1cm,
                                    draw=black,
                                    fill=gray!20]
  \tikzstyle{defensor_decision}=[rectangle,
                                    thick,
                                    minimum size=1cm,
                                    draw=black,
                                    fill=white]

  \node[uncertain](D)   {$\Theta $};
  \node[defensor_uncertain] (A) [left of=D]  {$D$};
  \node[decision] (B) [right of=D] {$A$};
  \node[utility]  (E) [below of=B] {$U_A$};

  \path (A) edge    node {} (D)
        (B) edge    node {} (D)
            edge    node {} (E)
        (D) edge    node {} (E)
        (A) edge[out=90, in=90, dashed] node {} (B);
\end{tikzpicture}
\caption{Defender analysis of Attacker problem.}\label{fig:adp}
\end{subfigure}
\caption{Influence diagrams for Defender and Attacker problems.}
\end{figure}

%The decision maker should consider games with incomplete information 
%{\color{black} 
\textcolor{black}{ As Example 1 shows, the solutions of games with complete information} 
might not be robust to perturbations in the
attacker's utilities and probabilities. Moreover, in many situations
the complete information assumption will not hold.
In both cases, the problem may be handled through an 
incomplete information game. The most common approach in such context uses BNE, see \cite{Hargreaves:2004} for details. %}
Alternatively, we use a decision analytic approach
based on ARA. % \citep{Rios-Insua:2009}. 
\textcolor{black}{ 
\cite{rios2012adversarial} and \cite{Banks2020}
discuss the differences between both concepts in, 
respectively, simultaneous and sequential games showing that they may lead to different solutions, with the interesting feature 
that ARA mitigates the common prior
assumption \citep{Antos} and provides prescriptive support to a decision maker.} %We describe the relation between both solution concepts in sequential games below.
%assuming common knowledge is not robust. Various models are proposed to handle cases of incomplete %information (\textcolor{red}{Reference}, imperfect information (\textcolor{red}{Reference} among %others. \textcolor{red}{They still suffer from common prior assumption-- discuss their potential %weakness in order to justify/motivate the use of ARA} Therefore, we perform a decision analytic %approach based on ARA 
%%%%%%%%%%%%%%%%%%%%%%%%%%%%%%%%%%%%
%Games with incomplete information can be addressed by performing a decision analytic approach based on ARA concepts, \cite{Banks:2015}. For this, we 
%
%\textcolor{red} {A general sentence about ARA: Bayesian game- its relation to standard game theory to give an idea about the big picture of its fit} 
%{\color{black} 
ARA facilitates the defender 
to acknowledge the uncertainty she might have
%} 
about $(u_A, p_A)$. Her problem 
is depicted in Figure~\ref{fig:ddp} as an influence diagram,
where $A$'s action appears as an uncertainty. % (\cite{shachter1986evaluating}). 
Her expected utility is %of defense $d$ would be
%
%\begin{equation*}
$\psi_D(d) = \int \psi_D(d,a)\, p_D(a \given d) \dd a $
%= \int \left[ \int u_D (d, \theta)\, p_D(\theta \given d,a) \dd \theta \right]\, p_D(a \given d) %\dd a ,$
%\end{equation*}
%
which %besides $p_D(\theta \given d,a)$ and $u_D(d, \theta)$, available from our discussion in %Section~\ref{sec:APS},
requires $p_D(a \given d)$, her assessment of the probability that the Attacker will choose $a$ after having observed $d$.
%To find $p_D(a|d)$, she could estimate the Attacker's utility function and his probabilities about both success $S$, conditional on $(a,d)$.
%Then, % we proceed as follows. 
Then, her optimal decision is $d^\opt_\text{ARA} = \argmax_{d \in \Dcal}\, \psi_D(d)$. 
\textcolor{black}{ Example 1 (cont.) below shows that it may be  
$d^\opt_\text{ARA} \neq d^\opt_\text{GT}$, due to the differences in informational assumptions.}
%that does not coincide with a Nash equilibrium.
%as it is based on different information and assumptions. %, see the example in Section 4. 

Eliciting $p_D(a \given d)$, with its %a 
strategic component,
is facilitated by analyzing $A$'s problem from $D$'s perspective (Figure~\ref{fig:adp}).
\textcolor{black}{ In order to accomplish it, the defender would use all information and judgment available to her} 
about $A$'s
utilities and probabilities.
However, instead of using point estimates for $u_A$ and $p_A$ 
\textcolor{black}{ to find
$A$'s best response $a^\opt(d)$ given $d$ (Section \ref{sec:APS}),} her uncertainty about the attacks 
would derive from %that about 
modeling $(u_A, p_A)$ through a distribution $F= (U_A, P_A) $ on the space of utilities and probabilities. With no 
loss of generality, assume that $U_A$ and $P_A$ are 
defined over a common probability space $(\Omega,{\cal A},{\cal P})$ with atomic elements $\omega \in \Omega$ \citep{Chung}.
%As an example of random probability measures, we consider a Dirichlet process $\cal{DP}(\alpha)$, introduced by Ferguson (1973), where the %parameter $\alpha$ is a positive measure. The Dirichlet process (DP) is a stochastic process on the space of the probability measures whose %realizations are (discrete a.s.) probability measures $P$, such that the expectation of any measurable subset is given by ${\cal E}[P(A)] = %\frac{\alpha(A)}{\alpha{\mathbb{R}}$. The latter property is very useful since the Defender might have an opinion on the Attacker's behavior and %she models it through the parameter $\alpha$, but relaxing the parametric assumption made earlier. We could go one step further, and we believe it %is appropriate in this context, by considering ${\cal P}_{A}$, a class of Dirichlet process (DP). In this case, we introduce an extra layer, more %realistic, about the Defender's knowledge and we might consider, for example, a class of DPs determined by a family of parameters $\alpha$ like in %Ruggeri (2010). Such class would model an imprecise knowledge by the Defender on the expected behavior by the Attacker and would allow for %simulations based on draws first of a DP in ${\cal P}_{A}$ and then of a probability measure from it,
This induces a distribution over the Attacker's expected utility $\psi_A(d,a)$, where the random expected utility %for %$A$
would be $\Psi _A^\omega  (d,a) = \int U_A^\omega  (a,\theta)
P_A^\omega (\theta \given d,a) \dd \theta$.
In turn, this induces a random optimal alternative defined through 
$A^{*}(d)^{\omega }= \argmax_{x \in \Acal} \Psi_A ^\omega (d, x)$.
Then, the Defender would find
$\label{plantio}
    p_D(a \given d) = \Pbb_F \left[ A^* (d) = a  \right]=
    {\cal P} \left\{ \omega : A^* (d)^{\omega }=a  \right\}
$ 
in the discrete case (and, similarly in the continuous one).
%Observe that $\omega$ and ${\cal P}$ 
%could be re-interpreted, respectively,
%as the type and the common prior in Harsanyi's doctrine.
%Then, $P_A ^\omega $ and $ U_A ^\omega$ respectively correspond to 
%$A$'s probability and utility given his  type, and
%$(d^*, \{ A ^* (d^*) ^{\omega}  \}) $ would constitute a BNE.
%Thus, in the sequential Defend-Attack game, we can operationally reinterpret the ARA approach in terms of 
%{\color{black}
%Harsanyi's BNE,
%}
 % although the underlying principles are different.
%Numerical enumeration can be used for low dimensional discrete cases. However, 
%\textcolor{red}{What about analytical solutions?--maybe discrete case-if none, then connect this to the upcoming subsection.}
Computationally, ARA models entail integration and optimization
procedures that can be challenging in many cases.
Therefore, we explore simulation based methods. 

%%%%%%%%%%%%%%%%%%%%%%%%%%%%%%%%%%%%%%%
\subsection{\textcolor{black}{ Monte Carlo based approach for ARA}} \label{sec:ARA-MC}

MC simulation approximates $p_D(a \given d)$ for each $d$, drawing $J$
samples $\left\lbrace \left( u_A^i, p_A^i \right) \right\rbrace_{i=1}^J $ from $F$ and making
$ \label{eq:MC}\notag
   \hat{p}_D(A\leq a \given d) = \frac{\# \lbrace A^* (d) \leq   a \rbrace}{J} %=  \argmax_{x \in \Acal} \, \Psi_A^i(x,d) \}}{J},
$ 
with $A^* (d) = \argmax_a  \int u_A^i(a,\theta) p_A^i (\theta \given d,a) \dd \theta$.
\textcolor{black}{ In case of multiple best attacker responses, ties are broken via SSE.}
%Thus, operationally, MC simulation estimates $p_{D}(A^*= a \given d)$,
%which
This is then used as input to 
the Defender's expected utility maximization, as reflected in Algorithm \ref{alg:mcmc_ara}, %{\color{black} 
which requires $\Dcal$ and 
$\Acal$ to be discrete (or discretized to the required precision
as shown in Section 4.2).
%}.

{\footnotesize
\begin{algorithm}[h!]
\linespread{0.7}\selectfont
\Input{$J$, $Q$, $P$}
\For{ all $d \in \Dcal$} {
	\For{\textcolor{black}{$j=1$} \KwTo $J$} {
		Sample $u_{A}^{j}(a, \theta) \sim U^{\omega}_{A} (a,\theta)$, 
		 $p_{A}^{j}(\theta \given d,a) \sim P^{\omega}_{A}(\theta \given d,a)$\;
    	\For{$a \in \Acal$} {
    	Generate samples $\theta_1, \dots, \theta_Q \sim p_{A}^{j}(\theta \given d,a)$\;
        Approximate $\widehat{\psi}^{j}_A(d, a) = \frac{1}{Q}  \textcolor{black}{\sum_i} u^j_A(a, \theta_i)$\;
   }
   Find {\color{black} $a_j^\opt(d)$, \textcolor{black}{via} %the element of interest of
   $\argmax_a \widehat{\psi}^j_A(d,a)$. 
   }
    }
	$\hat{p}_{D}( a \given d) = \frac{1}{J}\sum_{j=1}^{J}{I[a_j^\opt(d) = a]} $\;
}
\For{all $d \in \Dcal$} {
Generate samples $(\theta_1, a_1), \dots, (\theta_P, a_P) \sim p_D(\theta \given d, a) \hat{p}_{D}( a \given d) $ \;
Approximate $ \widehat{\psi}_D(d) = \frac{1}{P} \textcolor{black}{\sum_i} u_D(d, \theta_i)$\;
}
Compute $\widehat{d}^\opt_\text{ARA} = \argmax_d \widehat{\psi}_D(d)$\;
\caption{MC based approach to solve the \textcolor{black}{two-stage sequential defend-attack} ARA model}\label{alg:mcmc_ara}
\end{algorithm}
}

\noindent \textcolor{black}{ Algorithm 4 requires } $|\Dcal| \times \big[ J \times $  $ \left( |\Acal| \times Q + 2 \right) + 2P \big] $ samples where $Q$ and $P$ are the number of samples required to
respectively approximate  $\int u_{A}^{i}(a, \theta)\, p_{A}^{i}(\theta \given d,a ) \dd \theta $ and  $\int \int u_{D}(d,\theta)\, p_D (\theta \given d, a)\, 
\hat{p}_{D} ( a \given d) \dd \theta \dd a$ to the desired precision.
%\textcolor{red}{Does P reflect to two values; since we are solving a double integral? In the %proof, I used $P_1$ and $P_2$. Feel free to improve notation, and update Proof above comp. %complexity accordingly}
Convergence follows by applying a %from two 
%applications of a 
uniform version of the SLLN twice,  see SM Section 1.2 for  details. 
%Similarly, the optimal decisions for the expected utility estimates converge to the optimal %decisions almost surely as the sample size diverges to infinity (\cite{shao1989monte}). %\textcolor{red}{we can write these MC results as Propositions as well.}
%\textcolor{red}{Decide if the following is necessary or pollutes the message-and its link to the %case of mixed strategies::: When $d^{*}$ is not unique, $\widehat{d^{*}_P}$ may not converge to %the global optima due to existence of many limit points. However, $\widehat{d^{*}_P}$ can be used %in practice as a good decision as its expected utility still converges to the optimal expected %utility. (\cite{shao1989monte})}
In high dimensional or continuous cases, and when model uncertainty dominates, methods that automatically focus on high probability-high impact events, as APS does, would typically be faster and more robust. %(\cite{rios2012adversarial}).

\subsection{\textcolor{black}{ Augmented probability simulation for ARA}}\label{sec:aps_ara}
%Consider an APS approach to solve the ARA problem. % We first provide several %observations concerning the relevant augmented probability models and, then, outline an algorithm. 
APS solves the ARA model by constructing augmented distributions. % for 
%both agents' problems. 
To solve $A$'s decision problem, 
\textcolor{black} { an APS in the space of the Attacker's random utilities and probabilities
is constructed}. For a given $d$, 
the random augmented distribution built is
$\Pi_A^{\omega}(a, \theta \given d) \propto U_A^{\omega}(a, \theta ) P_A^{\omega}(\theta \given d, a)$, its marginal
$\Pi_A^{\omega}(a \given d) = \int \Pi_A^{\omega}(a, \theta \given d) \dd \theta$
 being proportional to $A$'s random expected utility $\Psi_A^{\omega}(d,a)$. Then, the random optimal
attack $A^*(d)^{\omega}$ coincides a.s.\ with the mode 
of the marginal $\Pi_A^{\omega}(a \given d)$. 
%of this random augmented distribution.
Consequently, by sampling $u_A(a, \theta) \sim U_A(a, \theta)$ and $p_A(\theta \given d, a) \sim P_A(\theta \given d, a)$, one can build $\pi_A(a, \theta \given d) \propto u_A(a, \theta) p_A(\theta \given d, a)$ which is a sample from $\Pi_A (a, \theta \given d)$. Then, $\text{mode}(\pi_A(a \given d))$ is a sample of $A^*(d)$, whose distribution is $\Pbb_F \left[ A^* (d) \leq a  \right] = 
p_D(A\leq a \given d)$, thus providing a mechanism to sample from such distribution. 
%%%%%%%%%%%%%%%%%%%%%%%%%%%%%%%%%%%%%%%%%%%%%%%%%%%%%%%%%%%%%%%
%%%%%%%  WHEN DO WE NEED THIS
%If necessary, we could estimate the value of $p_D(a \given d)$, drawing $J$ samples %from $p_D(a \given d)$ and counting frequencies as we did in \ref{sec:ARA-MC}.
%As we will see, moving to APS prevents us to compute $p_{D} (a \given d)$ for each $d \in \mathcal{D}$ (we just need to sample from this distribution), thus removing the dependence on $\given \mathcal{D} \given$ in the complexity.
{\color{black} 
As before, if $\pi_A(a \given d)$ has multiple global modes,
ties are broken via SSE to chose the mode of interest.}
Next, using backwards induction, an augmented distribution for the Defender's problem is introduced as $\pi_D(d, a, \theta) \propto u_D (d, \theta)~ p_D(\theta \given d, a)~p_D(a \given d)$. Its marginal $\pi_D(d)=\int\int \pi_D(d, a, \theta) \dd a  \dd \theta$ is proportional to the expected utility $\psi_D(d)$ and, consequently, $d^\opt_\text{ARA}  = \mode{(\pi_D(d))}$. Thus, one 
just needs to sample $(d, a, \theta )\sim \pi_D(d, a, \theta)$ 
and estimate its  mode in $d$.

\textcolor{black}{
Algorithm \ref{alg:ARA_APS_2} summarizes 
a nested MH based procedure for APS with states ($d$, $a$, $\theta$)
%be the current state of the Markov chain. The algorithm 
%samples a candidate defense  $\tilde{d}\,$  from a proposal generating distribution $g_D (\tilde{d} \given d)   $,
%a candidate $\tilde{a}$ from 
%$ p_A(a \given \tilde{a})$ using the 
%{\color{black}
%Attacker's APS
%} 
% explained before, and $\tilde{\theta} \sim p_D(\theta \given \tilde{d}, \tilde{a})$. 
%These samples are accepted with probability $\alpha = \min \left \lbrace 1, \frac{u_D(\tilde{d}, \tilde{\theta}) \cdot g_D (d \vert \tilde{d}) }{u_D(d, \theta ) \cdot g_D (\tilde{d} \given d)} \right \rbrace$. 
whose stationary distribution is 
$\pi_D(d, a, \theta)$.
%as reflected in 
Proposition \ref{conv_apsara}
%, providing
\textcolor{black}{ provides conditions for the a.s.\ convergence of the output of
the Algorithm to the optimal decision $d^\opt_\text{ARA}$.}
%\textcolor{red}{This is} similar to %those of
%Proposition 1. 
\textcolor{black}{Assumptions a), b), c), and d), are similar to those of
Proposition 1, hence are not repeated.}
}
{\small
\begin{algorithm}[h!]
\linespread{0.7}\selectfont
\SetKwFunction{Fatk}{sample\_attack}
\SetKwProg{Fn}{function}{:}{}

\Fn{\Fatk{$d$, $M$, $K$, $g_A$, $U_A$, $P_A$}}{
\Initialize{ $a^{(0)}$  }
Draw $u_A(a, \theta) \sim U_A(a,\theta)$\;
Draw $p_A(\theta \given d,a) \sim P_A(\theta \given d,a)$\;
Draw $\theta^{(0)} \sim p_A(\theta \given d, a^{(0)})$\;
\For(\Comment*[f]{Inner APS}){$i=1$ \KwTo $M$} {
	Propose new attack $\tilde{a} \sim g_A(\tilde{a} \given a^{(i-1)})$ \;
	Draw $\tilde{\theta} \sim p_A(\theta \given d, \tilde{a})$\;
    Evaluate acceptance probability
    $
    \alpha = \min \left \lbrace 1, \frac{u_A\left(\tilde{a}, \tilde{\theta}\right) \cdot g_A\left(a^{(i-1)} \vert \tilde{a} \right) }{u_A\left(a^{(i-1)}, \theta^{(i-1)} \right) \cdot g_A\left(\tilde{a} \given a^{(i-1)}\right)} \right \rbrace
    $\;
    With probability $\alpha$ set $(a^{(i)}  ,
    \theta^{(i)} )= ( \tilde{a},  \tilde{\theta})$. Otherwise, set $ (a^{(i)},\theta^{(i)})
    = ( a^{(i-1)}, \theta^{(i-1)}) $.\;
    }
%Discard first $$ samples and compute {\color{black} the mode of interest} of the rest of draws $\{a^{(i)}\}$\;
Discard first $K$ samples and \textcolor{black}{compute $a^\opt(d)$ based one the mode(s) of the remaining draws $\{a^{(K+1)},...,a^{(M)}\}$.}\;

%record {\color{red} the mode of %interest from 
%the rest of draws $\{a^{(K+1)},...,a^{(M)}\}$} as $a^\opt(d)$.}\;
\textbf{return} $a^\opt(d)$\;
}

\Input{$d$, $U_A$, $P_A$, $M$, $K$, $N$, $R$, $g_D$ and $g_A$ proposal distributions}
\Initialize{ $d^{(0)} = d$ }
Draw $a^{(0)} \sim p_A( a \given d^{(0)})$ using \Fatk{$d^{(0)}$, $M$, $K$, $g_A$, $U_A$, $P_A$}\;
Draw $\theta^{(0)} \sim p_D(\theta \given d^{(0)}, a^{(0)})$\;
\For(\Comment*[f]{Outer APS}){$i=1$ \KwTo $N$} {
    Propose new defense $\tilde{d} \sim g_D(\tilde{d} \given d^{(i-1)})$\;
    Draw $\tilde{a} \sim p_A( a \given \tilde{d})$ using \Fatk{$\tilde{d}$, $M$, $K$, $g_A$, $U_A$, $P_A$}\; 
    Draw $\tilde{\theta} \sim p_D(\theta \given \tilde{d}, \tilde{a} )$\;
    Evaluate acceptance probability
    $
    \alpha = \min \left \lbrace 1, \frac{u_D \left( \tilde{d}, \tilde{\theta} \right) \cdot g_D\left(d^{(i-1)} \vert \tilde{d} \right) }{u_D \left( d^{(i-1)}, \theta^{(i-1)} \right) \cdot g_D\left(\tilde{d} \given d^{(i-1)}\right) } \right \rbrace
    $\;
    With probability $\alpha$ set 
    $(d^{(i)} , a^{(i)}, \theta^{(i)})=(\tilde{d}, \tilde{a}, \tilde{\theta})$. 
    Otherwise, set $(d^{(i)}, a^{(i)}, \theta^{(i)})
    = ( d^{(i-1)}, a^{(i-1)}, \theta^{(i-1)})$.\;
    }
%Discard first $R$ samples and compute mode $\widehat{d}^\opt_\text{ARA}$ of rest of draws $\{d^{(i)}\}$\;
\textcolor{black}{Discard first $R$ samples and estimate mode from the rest of draws 
$\{d^{(R+1)},...,d^{(N)}\}$ as $\widehat{d}^\opt_\text{ARA}$}\;
Record %it as 
$\widehat{d}^\opt _\text{ARA}$.\;
\caption{MH based APS to approximate ARA solution in the \textcolor{black}{two-stage sequential defend-attack} game.} \label{alg:ARA_APS_2}
\end{algorithm}
}
%%%%%%%%%%%%%%%%%%%%%%%%%%%%%%%%%%%%%%%%%%%%%
%\begin{prop}\label{prop:conv_nested}
%If the Attacker's and Defender's utility functions are positive and integrable, %$p_A(\theta \given d,a), p_d(\theta \given d,a) > 0$ $\forall a, \theta$ and %$\mathcal{A}$, $\mathcal{D}$, $\Theta$ are either discrete or discretized %continuous variables represented with intervals in $\mathbb{R}^n$, , the output of %Algorithm \ref{alg:MHdefenderAPS2} defines a Markov Chain with stationary %distribution $\pi_D(d, \theta, a^\opt(d))$. The mode of the marginal samples of %$d$ from the Markov chain coincides with $d^\opt_\text{GT}$.
% its marginal distribution on the Defender's decision space, $\pi_D(d)$, coincides %with $d^\opt_\text{GT}$. 
%\end{prop}
%%%%%%%%%%%%%%%%%%%%%%%%%%%%%%%%%%%%%%%%%%%%%%%%%
%%%%%%%%%%%%%%%%%%%%%%%%%%%%%%%%%%%%%%%%%%%%%%%%%%%%
%\begin{comment}

\textcolor{black}{
\begin{prop}\label{conv_apsara}
{\color{black} Assume that $u_D$ and the utilities in the support of $U_A$ (a.s.) satisfy 
condition a); %are positive and continuous in, respectively, $(d,\theta) $ and $(a,\theta )$; 
$\mathcal{A}$ and $\mathcal{D}$ 
satisfy condition b); %are %either discrete or
%discretized continuous variables represented with 
%compact sets in $\mathbb{R}^n$; 
$p_D(\theta \given d,a)$ and the distributions in the support of $P_A(\theta \given d,a)$ (a.s.) satisfy condition c); %are positive and 
%continuous in $d$ and $a$; 
%the products of utilities and probabilities are integrable; 
the proposal generating distributions $g_A$ and $g_D$,
satisfy condition d);} and the Attacker's best response set is finite for every $(u_A, p_A) \in U_A \times P_A$. 
Then, Algorithm \ref{alg:ARA_APS_2} defines a 
Markov chain with stationary distribution $\pi_D(d, \theta, a)$. 
Moreover, a consistent mode estimator based 
on the marginal samples of $d$ from this Markov chain a.s. approximates $d^\opt_\text{ARA}$.
\end{prop}
%\vspace{-0.1in}
}
%\end{comment}
\begin{comment}
\textcolor{black}{
\begin{prop}\label{conv_apsara}
{\color{black} Assume} that $u_D$ and the utilities in the support of $U_A$ (a.s.) 
%satisfy condition a); 
{\color{black} are positive and continuous in, respectively, $(d,\theta) $ and $(a,\theta )$};
$\mathcal{A}$ and $\mathcal{D}$ 
%satisfy condition b); %are %either discrete or
%discretized continuous variables represented with 
{\color{black} are compact sets}; %in $\mathbb{R}^n$; 
$p_D(\theta \given d,a)$ and the distributions in the support of $P_A(\theta \given d,a)$ (a.s.) %satisfy conditions c) and d); 
{\color{black}are positive and continuous in $d$ and $a$}; 
%the products of utilities and probabilities are integrable; 
the proposal generating distributions $g_A$ and $g_D$,
%satisfy condition d)
{\color{black}have support $\mathcal{A}$ and $\mathcal{D}$}; and the Attacker's best response set is finite for every $(u_A, p_A) \in U_A \times P_A$. 
Then, Algorithm \ref{alg:ARA_APS_2} defines a 
Markov chain with stationary distribution $\pi_D(d, \theta, a)$. 
Moreover, a consistent mode estimator based 
on the marginal samples of $d$ from this Markov chain a.s. approximates $d^\opt_\text{ARA}$.
\end{prop}
%\vspace{-0.1in}
}
\end{comment}

\noindent Algorithm \ref{alg:ARA_APS_2} requires generating $N \times \left( 2M + 5 \right) + 2M + 4$ samples from multivariate distributions, in addition to the cost of the convergence checks and mode computations, where $M$ and $N$ are \textcolor{black}{the number of MCMC iterations for the attacker and defender's APS, respectively. We emphasize that} this algorithm removes the need for loops over ${\cal A}$ and ${\cal D}$, and can be  applied to continuous decision sets without discretizing.
%To the best of our knowledge, this is the first ARA solution method that can be used directly with continuous decision sets.
%This would be an excellent choice when facing a problem where the cardinality of these spaces is large or decisions are continuous. 
%{\color{black}
\textcolor{black}{Finally, note that} it may be beneficial
to combine MC and APS in \textcolor{black}{some discrete cases. For instance,} if $|{\cal D}|$ is low, it 
is convenient to \textcolor{black}{use MC} to estimate $p_D(a \given d)$ for each $d$ by drawing $J$ samples $a \sim p_D(a \given d)$ and counting frequencies as in
Section \ref{sec:ARA-MC}. %Then, in the Defender's APS, for samples from $a \sim p_D(a \given d)$, instead of invoking the attacker's APS, direct sampling from the estimate $\hat{p}_D(a \given d)$ \textcolor{red}{could be more efficient}.
\textcolor{black}{Using direct samples from this estimate,  $\hat{p}_D(a \given d)$,
instead of invoking the attacker's APS could be more efficient for retrieving samples from $a \sim p_D(a \given d)$ within the Defender's APS.}

%}
% Finally, \ref{ec:gibbs_ara} provides a Gibbs based algorithm.

%%%%%%%%%%%%%%%%%%%%%%%%
\paragraph{Example 1 (Cont.)}
\label{subsec:ara_approach}
Complete information is %assumed not to be available any more. 
no longer available.
$D$'s beliefs over
$A$'s judgments are described through $P_A$ and $U_A$.
Assume 
%the Attacker knows the Defender's decision,
$A$'s random probability of success is modeled as 
$P_A(\theta  = 1 \given d,  a=1) \sim Beta (\alpha_d, \beta_d)$ with
 $\alpha_d$  and $\beta_d$  in Table \ref{tab:dist}
 \textcolor{black}{ 
(with $p_D(\theta =1 \given d,a)$ in  
Table \ref{tab:prob} as expected values)}. 
In addition, $A$'s  risk coefficient $e$ is uncertain,
with $e \sim {\cal U} (0,2)$, 
inducing the random utility $U_A (c_{A})$.
\begin{figure}[h!]
\begin{subfigure}{0.35\textwidth}
\centering
\includegraphics[width=\textwidth]{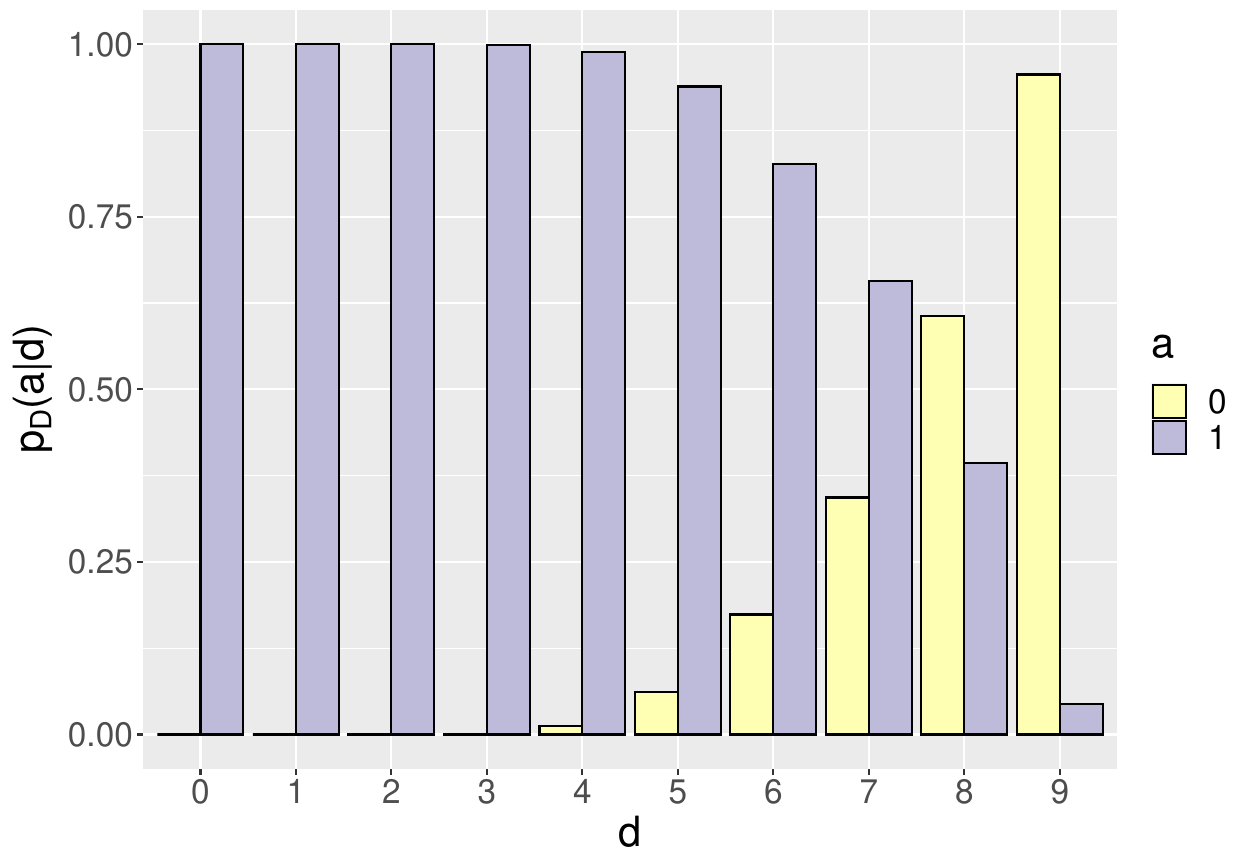}
\caption{MC estimation of $p_D(a \given d)$}\label{fig:MC-prob-ARA}
\end{subfigure}
\hfill
\begin{subfigure}{0.35\textwidth}
\centering
\includegraphics[width=\textwidth]{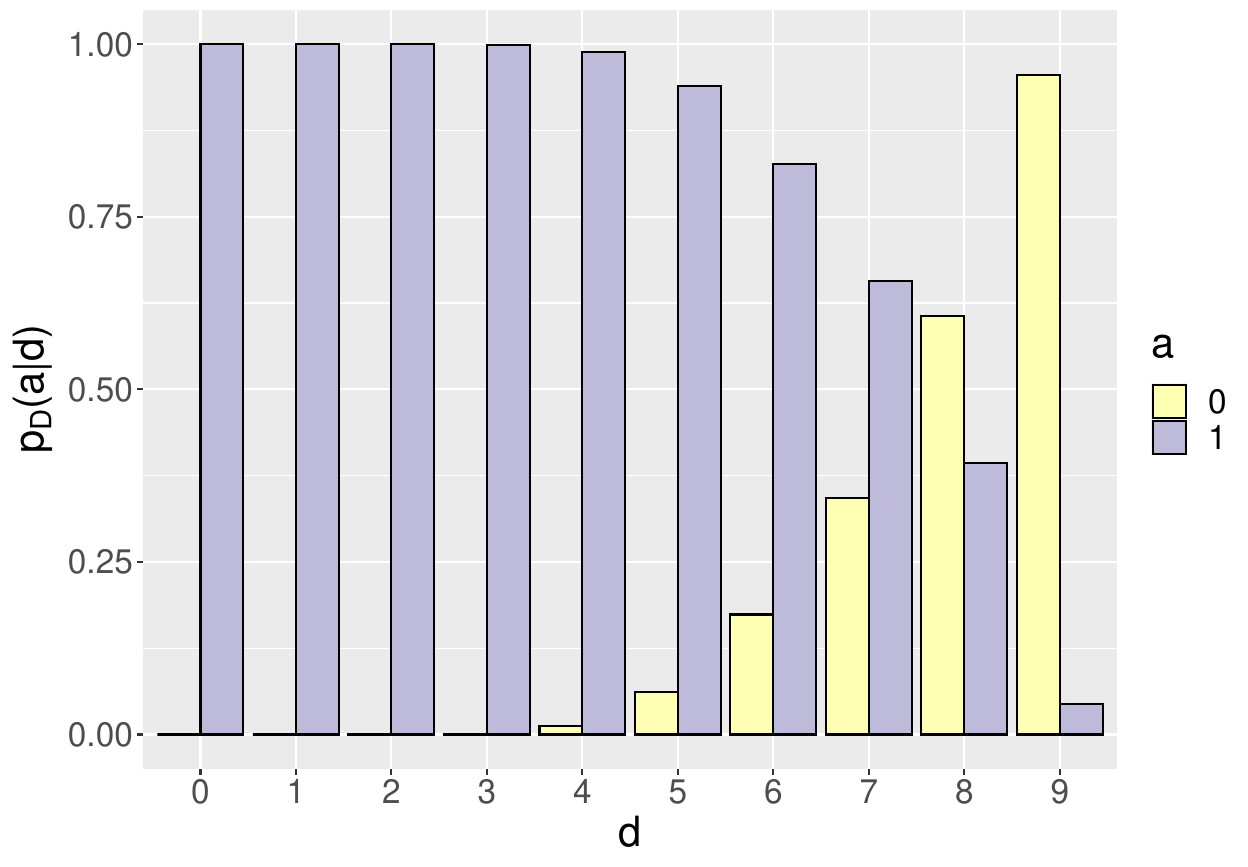}
\caption{APS estimation of $p_D(a \given d)$}\label{fig:APS-prob-ARA}
\end{subfigure}
\caption{Estimation of $p_D(a \given d)$ through ARA}\label{fig:prob-ARA}
\end{figure}
\begin{figure}[h!]
\begin{subfigure}{0.35\textwidth}
\centering
\includegraphics[width=\textwidth]{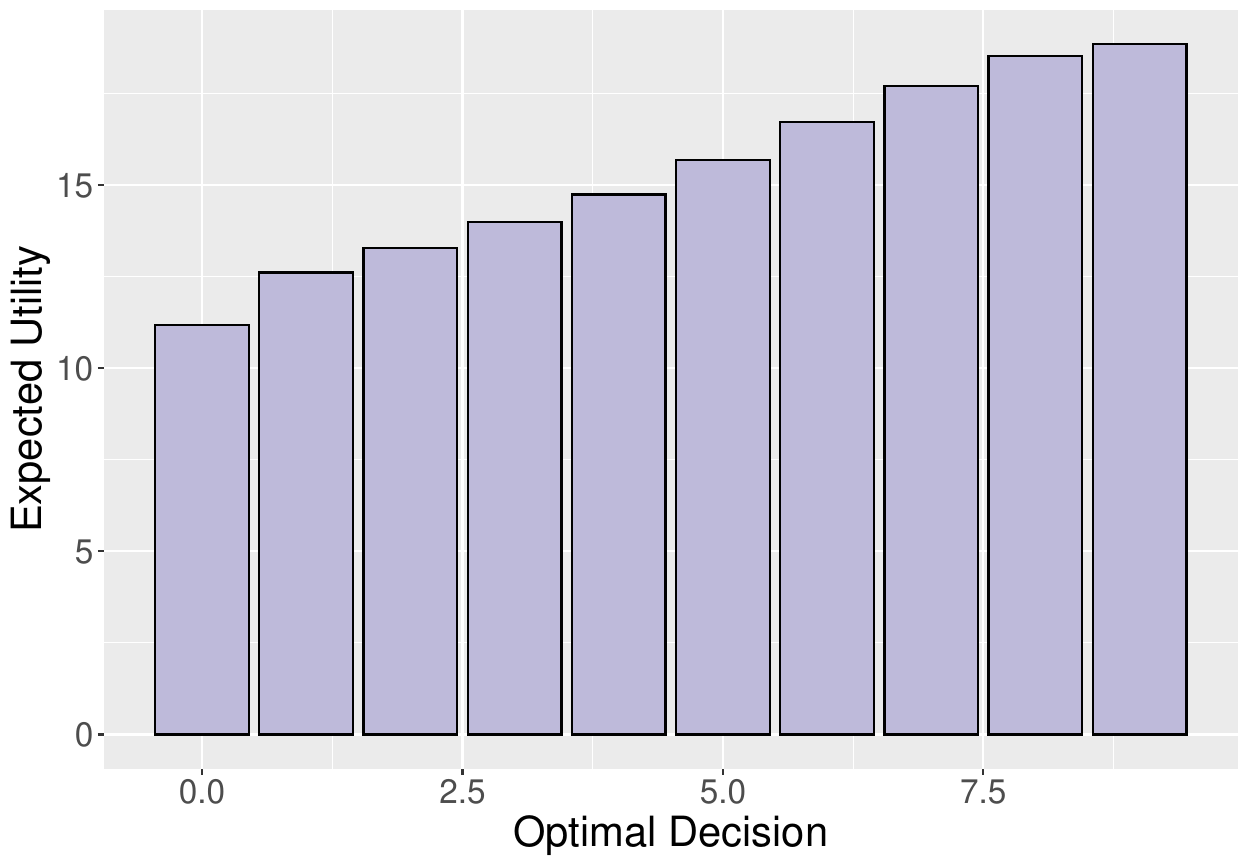}
\caption{MC solution}\label{fig:MC-ARA}
\end{subfigure}
\hfill
\begin{subfigure}{0.35\textwidth}
\centering
\includegraphics[width=\textwidth]{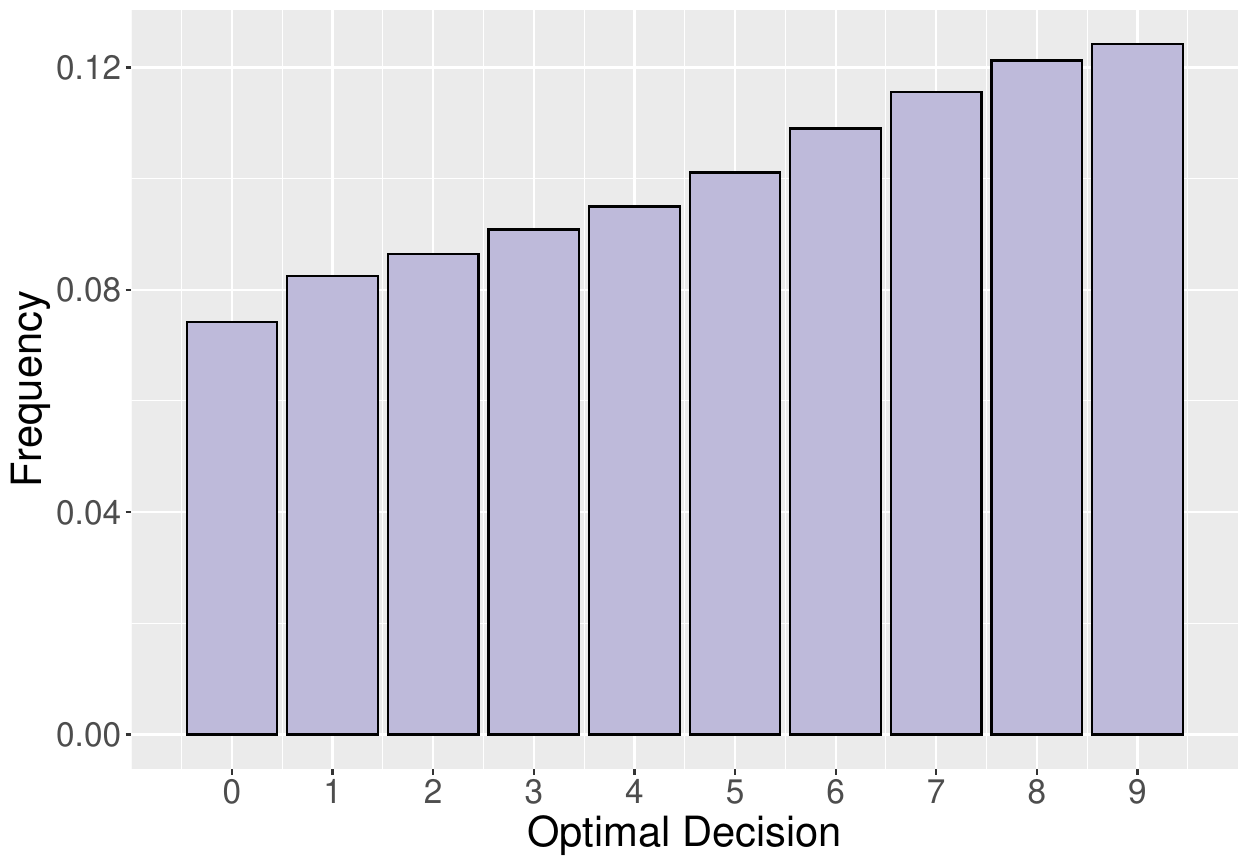}
\caption{APS solution}\label{fig:APS-ARA}
\end{subfigure}
\caption{ARA solutions for the Defender}\label{fig:ARA}
\end{figure}
In this case, for APS, as the cardinality of ${\cal D}$ 
is small, one can estimate the value of $p_D(a \given d)$ for each $d$. Figure \ref{fig:prob-ARA} presents the estimates
$\hat {p}_D(a \given d)$, %over attacks $a$, given defense $d$,
obtained using MC and APS, 
%{\color{black} 
 which appear quite similar: the maximum absolute difference between MC and APS estimates is less than $0.01$.
%}
Next, the ARA solution is computed. Figure \ref{fig:MC-ARA} shows the MC estimation
of her expected utility;  Figure \ref{fig:APS-ARA} presents the frequency of samples from the marginal $\pi_D(d)$. Its mode 
coincides with the optimal defense, $d_\text{ARA}^* = 9$, in agreement with the MC solution. \textcolor{black}{ Observe that in this case the ARA decision $d_\text{ARA}^*$ differs from 
 the game-theoretic solution $d_\text{GT}^*$,
the informational assumptions being different: the ARA decision
appears to be more conservative, as it suggests a safer but more expensive defense. SM Section 4.2} replicates 
the experiment dividing $\alpha_d$ and $\beta_d$ by 100 in Table \ref{tab:dist}, thus inducing more uncertainty about the attacker's probabilities. The ARA solution ($d^*_{ARA}=9$) remains stable despite these changes.
\hfill $\triangle$
%Of course, as in any decision analysis, we could perform sensitivity analysis with, e.g.\ alternative $c$ values, say between $0.1$ and $1$.
 %However, we shall not focus on that, see \cite{rios2012adversarial} for a relevant analysis.
%\%pagebreak

%%%%%%%%%%%%%%%%%%%%%%%%%%%%%%%%%%%%%%%%%%%%
%\subsubsection{Robustness of the ARA solution}
%Values of distributions for $P_A$ and $c$

%Table with results\\
%Discussion\\
%Link to minimizing the maximum regret\\

%\newpage
%%%%%%%%%%%%%%%%%%%%%%%%%%%%%%%%%%%%%%%%%%%%%%%%%%%%%%%%%%%

%%%%%%%%%%%%%%%%%%%%%%
\subsection{Sensitivity analysis of the ARA solution}
The ARA approach leads to a decision analysis problem with the peculiarity of including a sampling procedure to forecast $A$'s actions. %To do so, we formulate his decision making problem and propagate our uncertainty about the adversary judgments to obtain the random optimal %adversarial action. 
A sensitivity analysis should  be conducted with respect to its inputs 
$(u_D (d,\theta ),\, p_D(\theta \given d,a),\, p_D(a \given d))$,
\textcolor{black}{ with focus on  
$p_D(a \given d)$, the most 
contentious one} as it comes from adversarial calculations based
on the random ingredients $U_A$ and 
 $P_A $.
We would proceed as has been done in Section \ref{sec:sens_anal} evaluating 
the impact of the imprecision on $U$ and $P$ over 
the attained expected utility %$d^{*UP}_{ARA}$ 
$\psi(d^{*UP}_\text{ARA})$ using classes $\mathscr{U}_A$, $\mathscr{P}_A$ of random utilities and probabilities; 
for each feasible $(U, P)$,
%from such classes,
  $p_D ^{UP} (a \given d) $ would be 
obtained to compute $d^{*UP}_\text{ARA}$,
estimating then the  maximum regret.

\section{Computational assessment}\label{sec:time_comp}
This section discusses computational complexity results, % of the proposed algorithms, %{\color{black} 
emphasizing \textcolor{black}{ the advantages 
of APS in  large scale game-theoretic settings.}
%}
  %, illustrating then the points with an 
%example.
%%%%%%%%%%%%%%%%%%%%%%%%%%%%%%%%%%%%%%%%%%%%%%%%%%%%%%%%%%%%%%%%%%%%
\subsection{Computational complexity}
Table \ref{tab:complexity} summarizes the computational complexity of MC and 
APS for solving games with complete and incomplete information.
%compiled from
%earlier sections.
Recall that  parameters $P$, $Q$, $N$ and $M$ 
would typically depend on the desired precision, as outlined in SM 
(Section 1.1).
{\small
\begin{table}[htbp]%{\textwidth}
\centering
\setlength\extrarowheight{-3pt}
\begin{tabular}{lcccc}
\toprule
& &   MC &  &   APS \\
\midrule
Complete  & & $|\Dcal| \times \left(|\Acal|\times Q + P \right)$ &  &
%$2\, \Big(|\Dcal|\times M + N \Big)$ 
$ N \times \left(2 M + 3 \right) + 2M + 2$ \\
Incomplete  & &  $|\Dcal| \times \left[ J \times \left( |\Acal| \times Q + 2 \right) + 2P \right] $ & & $N \times \left( 2M + 5 \right) + 2M + 4$ \\
\bottomrule
\end{tabular}
\caption{MC and APS required sample sizes for games with complete and incomplete information} \label{tab:complexity}
\end{table} 
}

%\textcolor{red}{if we decide to keep old Alg 2-Gibbs-mention its potential advantage-adding dependence on defenders decision space vs MCMC iterations}
%\noindent In the case of complete information, Algorithm \ref{alg:MHdefenderAPS2} can be used to remove dependence of the complexity on the cardinality of the Defender's decision space as well. 
 
\noindent \textcolor{black}{ Continuous decision spaces need to be discretized to approximate the MC solution, as Section \ref{sec:computational_comp} illustrates.} 
The discretization step impacts the solution precision and the
cardinalities of $\Dcal$ and $\Acal$  which, in turn, affect complexity. %of both algorithms.
In contrast, APS does not discretize the decision sets.
\textcolor{black}{ As Table \ref{tab:complexity} shows,}
 the number of MC samples depends on the cardinality of the Defender's and Attacker's decision spaces, whereas 
this dependence is not present in APS.  Thus, this approach  
would be expected to be more efficient than MC for problems with 
large decision spaces as showcased \textcolor{black}{in the following subsection}.  
%In particular, the cardinalities $|\Dcal|$ and $|\Acal|$ do not affect the complexity of APS %througs Algorithms  * and 
%\ref{alg:MHdefenderAPS2}.%, while this is not possible for MC

%%%%%%%%%%%%%%%%%%%%%%%%%%%%%%%%%%%%%%%
\subsection{A computational comparison} \label{sec:computational_comp}
A simple game with continuous decision spaces \textcolor{black}{ serves to compare the scalability of %{\color{black} 
the proposed approaches.}
%}
\textcolor{black}{ The agents make their respective decisions $d, a \in [0,1]$ 
concerning the proportion of resources  invested
to defend and attack a server whose value is $s$.} Let $\theta$ be the proportion of losses for the defender under a successful attack,  % is denoted as , %The value of the server is $s$.
modeled with a Beta distribution with parameters $\alpha(d,a)$ and $\beta(d,a)$, with $\alpha$ ($\beta$) increasing in $a$ ($d$) and decreasing in $d$ ($a$).
$D$'s payoff is $f(d, \theta) = (1 - \theta)\times s - c\times d$, where $c$ denotes her unit resource cost. She is constant risk averse
with utility strategically equivalent to $1 - \exp \left( -h\times f(d,\theta) \right)$, $h > 0$. $A$'s payoff is $g(a, \theta) = \theta \times s - e \times a$, where $e$ denotes $A$'s unit resource cost. He is constant risk prone with utility strategically equivalent to $\exp \left( -k\times g(a,\theta) \right)$, $k > 0$.

%\begin{verbatim}
%    In the EC there is a proof of convergence of H copies for ARA. 
%    But here we use H copies for games.
%    Also in the proof we should distinguish more clearly the 2 cases:
%    1) H is fixed, in this case the convergence follows from standard MCMC results.
%    2) H varies according to some cooling scheme. The proof is the one
%    from Muller.
%\end{verbatim}

\begin{figure}[htbp]
\centering
\includegraphics[width=0.35\textwidth]{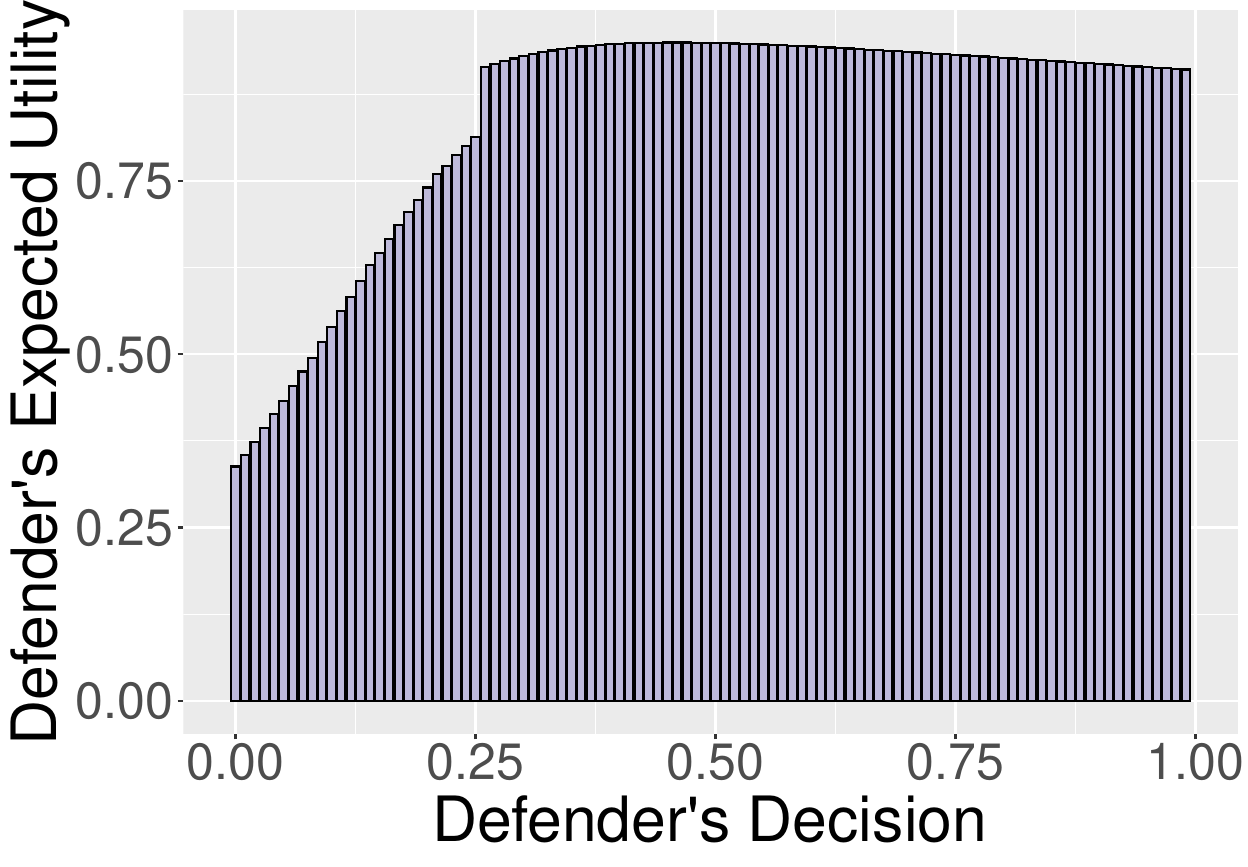}
\caption{Defender's expected utility surface. Optimal decision in yellow.}\label{fig:DEU}
\end{figure}

Figure \ref{fig:DEU} provides MC estimates of $D$'s expected utility, which is arguably flat \textcolor{black}{ around the optimal defense}. To increase efficiency of APS
in detecting the mode, we replace the marginal augmented distribution $\pi_D(d \given a^\opt(d))$ by its power transformation $\pi_{D}^{H}(d \given a^\opt(d))$ (with $H$ 
designated {\em augmentation parameter}) to make the distribution more peaked around the mode \citep{M:1999}. %, and 
 %\citep{muller2004optimal}, %his idea is related to simulated annealing (\cite{KGV:1983}), 
%(see %\cite{EkinAktekin:2016}, \cite{Ekin:2018} and 
%see \cite{ekin2021decision} for an application.
This underlines another advantage of APS over MC, as it improves the efficiency of direct MC sampling from flat regions. SM Section 2.2 presents details and sketches a convergence proof.
\textcolor{black}{
Sampling from $\pi_{D}^{H}(d \given a^\opt(d))$ can be performed using Algorithm \ref{alg:MHdefenderAPS2} 
drawing $H$ copies of $\tilde{\theta}$, instead of just one. %\citep{muller2004optimal}.
The acceptance probability for $D$'s problem at a given
iteration would be  
$     \min \left \lbrace 1, \prod_{t=1}^H \frac{u_D(\tilde{d}, \tilde{\theta}_t)}{u_D(d^{(i-1)}, \theta^{(i-1)}_t)}\right \rbrace,
$ 
(Section 2.2).} $A$'s problem includes a similar augmentation. The  augmentation parameters for $A$ and $D$ will be
designated inner and outer powers, respectively.
  %However, they 
 %would still need to be optimized. %\cite{tierney1994markov} presents the MCMC convergence of $\pi_{D}^{H}(d)$ in number of MCMC %iterations for a fixed value of augmentation parameter, $H$. 
 %They are chosen with a diagnostic which increases the power based on an initial exploratory simulation from many independent sequences, 
%see \cite{gelman1992inference}.
%Then, it is enough to sample from the joint density and compute the mode of the marginal draws along the Markov chain for a value of $H$ on a given schedule. %(\cite{JJP:2007}). %Practical convergence in $H$ could be assessed for reasonable values of $H$ such as $ J=5,10,25,50,100$ in most applications.
%The augmentation parameter could be utilized to handle very flat expected utility surfaces. 
%using $J$ \textit{iid} copies of the random vector $(\xi_1,\xi_2)$, we can write down the augmented distribution, 
%$$\pi_{J} (x,  \bm{\xi}_{1J} , \bm{\xi}_{2J}) \propto \displaystyle\prod_{j=1}^J\ %(-cx + s(\xi_{2j}) y^{*}_{1j} +r y^{*}_{2j})p(\xi_{1j},\xi_{2j}) , $$
%where 
%$\xi_{i,j}$ is the $j^{th}$ draw of $\xi_{i}$, $\bm{\xi}_{iJ}= \{ \xi_{i,1},\ldots, %\xi_{i,J} \}$ for $i=1,2$, $\bm{\xi}_{j}= \{ \xi_{1j}, \xi_{2j} \}$, %$y^{*}_{1j}=y_{1}^{*}(x,\bm{\xi}_j)$ and $y^{*}_{2j}=y_{2}^{*}(x,\bm{\xi}_j)$. 
%\textcolor{red}{Should we include the mathematical representation of APS to the power of H with H copies of random variables: Or would it confuse the reader.}
%DECIDE ON THE ABOVE:

%{\color{black} 
To solve this problem with MC, the decision sets must be discretized. The chosen discretization step will affect the solution precision as well as the cardinality of the (discretized) decision spaces, thus affecting MC performance. For instance, to get a solution with precision $0.1$, we discretize $\Dcal$ through $d= \{0, 0.1,0.2,...,1\} \in \Dcal$, and the \textcolor{black}{ cardinalities of the decision spaces are}  $|\Dcal| = |\Acal| = 11$. With precision $0.01$, $|\Dcal| = |\Acal| = 101$, and so on. In contrast, APS does not require discretizing. To deal with continuous decision sets, all we need is to use continuous proposals $g_D$ and $g_A$ in Algorithm \ref{alg:ARA_APS_2}. In this example, we use uniform distributions. 
%Clearly, the higher the precision required in the solution, the higher the sampling effort, as the mode of the corresponding marginal augmented distributions needs to be better resolved.
%}a

\textcolor{black}{ Let us investigate the trade-off between precision and sample size, finding a threshold beyond which APS (MC) is faster for
greater (smaller) precision.} To do so, the minimum number of required MC and APS samples to achieve an optimal decision within a given precision are computed for both 
$A$ and $D$ problems, respectively designated as inner and outer samples. 
In addition, for APS,  we also report the %minimum 
{\em inner} and {\em outer} powers used to achieve that precision.
For fair comparison, this study conducts several parallel replications of MC and APS with an increasing number of samples. When $90 \%$ of the solutions coincide with the optimal decision (computed with MC using a large number of samples) with the required precision, the algorithm is declared to have \textit{converged} for such number of iterations. 
%{\color{black} 
Using this number of iterations, the time employed to compute the optimal decision is recorded.
%}

%A fair comparison of MC and APS running times is provided. 

%
%\begin{table}
%\centering
%\begin{tabular}{cccc}
%\toprule
%Precision & Outer Iterations & Inner Iterations & Time \\
%\midrule
%0.1  &   1000 &  1000 & $0.020 \pm 0.001$ \\
%0.01 &   111000 &  111000 & $259.9 \pm 37.9$ \\
%\bottomrule
%\end{tabular}
%\caption{MC times and optimal iterations for different precisions.\ATB{juntaria las dos tablas en una, puse ejemplo}}\label{tab:MC_time}
%
%\end{table}
%%%%%%%%%%%%%%%%%%%%%%%%%%%%%%%%%%%%%%%%%%%%%%%%%%%%%%%%%%%
%\begin{table}
%\centering
%\begin{tabular}{cccccc}
%\toprule
%Precision & Outer Iterations & Inner Iterations & Outer Power & Inner Power & Time %\\
%\midrule
%0.1  &   50   &  50  & 40     & 40   & $0.039 \pm 0.001$ \\
%0.01 &   1000 &  100 & 100000 & 1000 & $74.9 \pm 12.6$ \\
%\bottomrule
%\end{tabular}
%\caption{APS times and optimal iterations and power for different precisions.}\label{tab:APS_time}
%
%\end{table}
%
{\small 
\begin{table}[htbp]
\centering
\setlength\extrarowheight{-3pt}
\begin{tabular}{llrrrrr}
\toprule
&       & \multicolumn{2}{c}{Samples} & \multicolumn{2}{c}{Power} & \\
\cmidrule(lr){3-4}\cmidrule(lr){5-6}
Precision & Algorithm & Outer & Inner & Outer & Inner & {Time (s)} \\
\midrule
0.1  & MC   &   1000 &  100 &    - &    - &  0.007 \\
     & APS  &     60 &  100 &  900 &   20 &  0.240 \\  [0.5 em]
0.01 & MC   & 717000 &  100 &    - &    - & 13.479 \\
     & APS  &    300 &  100 & 6000 &  100 &  2.461 \\
\bottomrule
\end{tabular}
\caption{Computational time, minimum number of required MC and APS samples and augmentation parameters at optimality for different precisions}
\label{tab:APS_time}
\end{table}
}

%\textcolor{red}{What is optimality and convergence? make it more concise-}
\textcolor{black}{ Table \ref{tab:APS_time} presents MC and APS  
computational times  
for precisions $0.1$ and $0.01$,} using a server node with 16 cores Intel(R) Xeon(R) CPU E5-2640 v3 @ 2.60GHz.
Computational runs show the need for only $1,000$ and $100$ MC samples
to, respectively, reach optimality to solve $D$ and $A$ problems with precision $0.1$: 
MC outperforms APS.
However, the performance of MC diminishes for higher precision. 
\textcolor{black}{ For instance, for precision $0.01$, %$|\Dcal| = |\Acal| = 101$ for MC and it becomes more demanding: 
APS outperforms MC.}  %as shown in Section \ref{sec:GT-APS}. 
Indeed, for MC, there is a factor of 200 between the time needed to obtain the solutions with precision 0.01 and 0.1. For APS, this factor is just 10, suggesting that it scales much better with precision. For smaller precision, such as $0.001$, we could not get a stable solution using MC even with a large number of samples ($P=10$M, $Q=100$k). 
Finally, 
 Figure \ref{fig:DEU} %observe 
 suggests that as the expected utility is flat around the optimal decision,
%{\color{black} 
%it is clear that 
MC will require a  high number of samples to resolve the maximum. For cases like this, APS is even more convenient,
as it enables sampling from the transformed marginal augmented distribution, more peaked around the mode than the 
initial distribution.
%}
%To sum up, in problems with large or continuous decision spaces and/or flat expected utilities, APS would be preferred over MC for its scalability.
%For such problems, MC could be used for an initial broad exploration of the decision space to determine regions of interest and, then, APS could be utilized to explore for the optimal decisions.

%%%%%%%%%%%%%%%%%%%%%%%%%%%%%%%%%%%%%%%%%%%%%%%%%%%%%%%
\section{A cybersecurity application}\label{sec:example2}
\begin{figure*}[h]
	\begin{center}
		\includegraphics[scale=0.45]{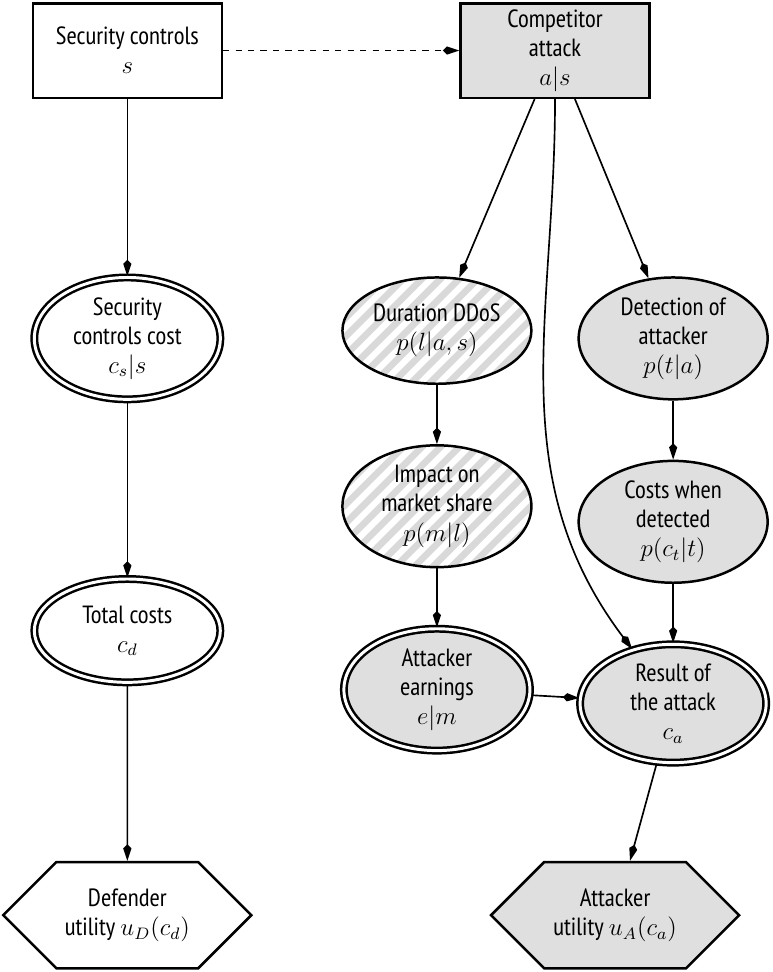}
		\caption{Bi-agent influence diagram %(BAID) 
		of the cybersecurity application.}
		\label{kakita}
	\end{center}
\end{figure*}

We %end up with 
consider a cybersecurity case
which illustrates
the proposed framework in a real world setting. %while showcasing the complexities of applications.
\textcolor{black}{ It simplifies the example  
in \cite{insua2019adversarial} by retaining only the adversarial cyber threat. 
The complexities of applications are apparent when observing the bi-agent influence diagram in 
Figure \ref{kakita}, which has  
additional uncertainty nodes beyond those in our 
template in Figure \ref{fig:tpsdg}.} 
%In particular, observe the additional 
%uncertainty nodes in the bi-agent influence diagram in Figure \ref{kakita} when compared with the template in Figure 
%\ref{fig:tpsdg}. 
\textcolor{black}{
An organisation ($D$) faces a competitor ($A$) that may attempt distributed denial of service (DDoS) attacks to undermine $D$'s site availability and
compromise her customer services.}

\textcolor{black}{
\noindent $D$ has to determine the subscription level to a monthly cloud-based DDoS protection system,
with choices $0$ (not subscribing), $5, 10,  \dots, 190,$ and $195$ gbps. 
$A$ must decide the intensity of his DDoS attack, viewed as the number of days (from 0 to 30) that he will attempt to launch it.
The duration of the DDoS may impact $D$'s market share,
due to reputational loss. %As in  \cite{insua2019adversarial}, Assume that all 
%The share is lost at a linear rate %$r$
%until it disappears after a few days of service unavailability.
%Being the sole competitor, 
%$A$ 
$A$ gains %$e$
%in terms of market share is $ e = m $, 
all market share lost by $D$,
%This 
which determines his earnings.
% The Attacker's earnings depend on the gained market share. 
However, he runs the risk of being detected with significant costs. Both agents aim at maximizing expected utility.
Details of the models at various nodes are in SM Section 5.} 

\begin{figure}[htbp]
\begin{subfigure}{0.4\textwidth}
\centering
\includegraphics[width=\textwidth]{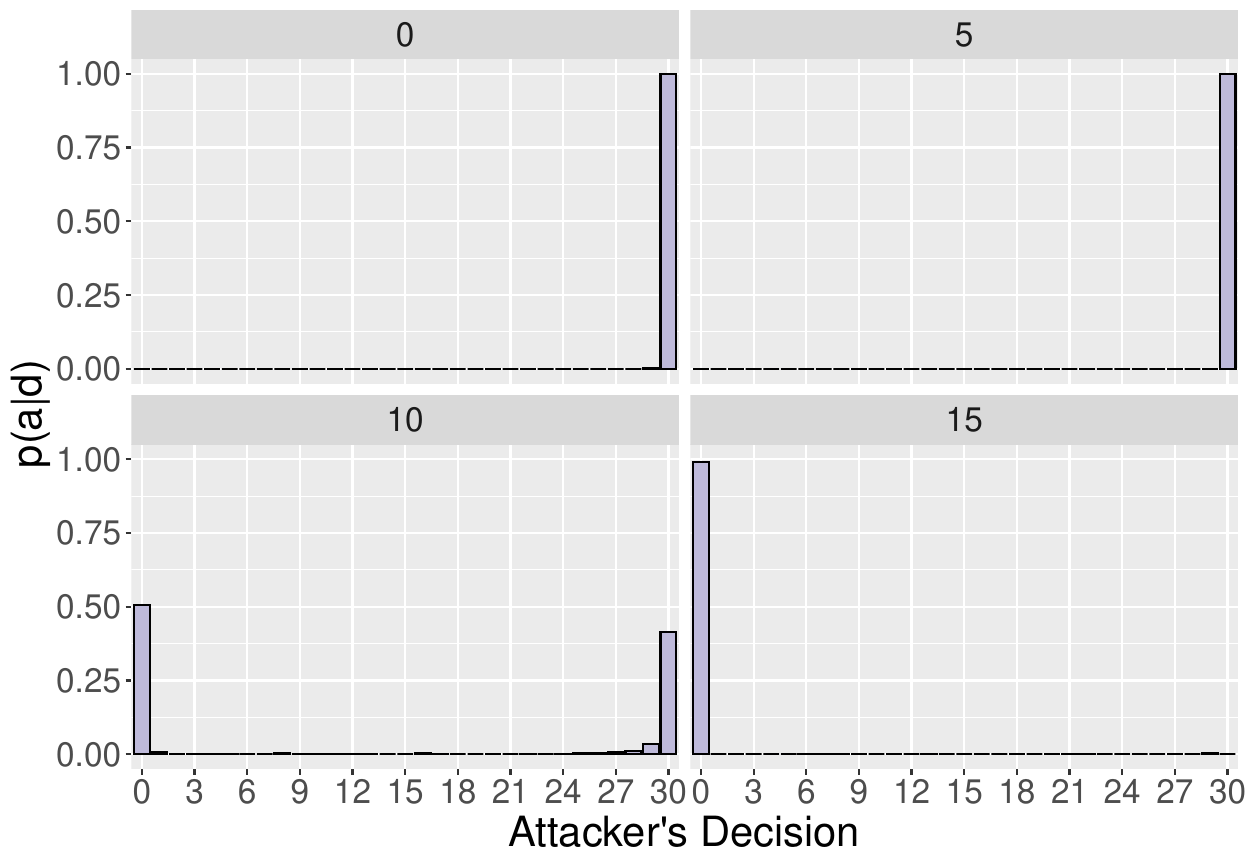}
\caption{Attack intensity for each decision}\label{fig:pad}
\end{subfigure}
\hfill
\begin{subfigure}{0.4\textwidth}
\centering
\includegraphics[width=\textwidth]{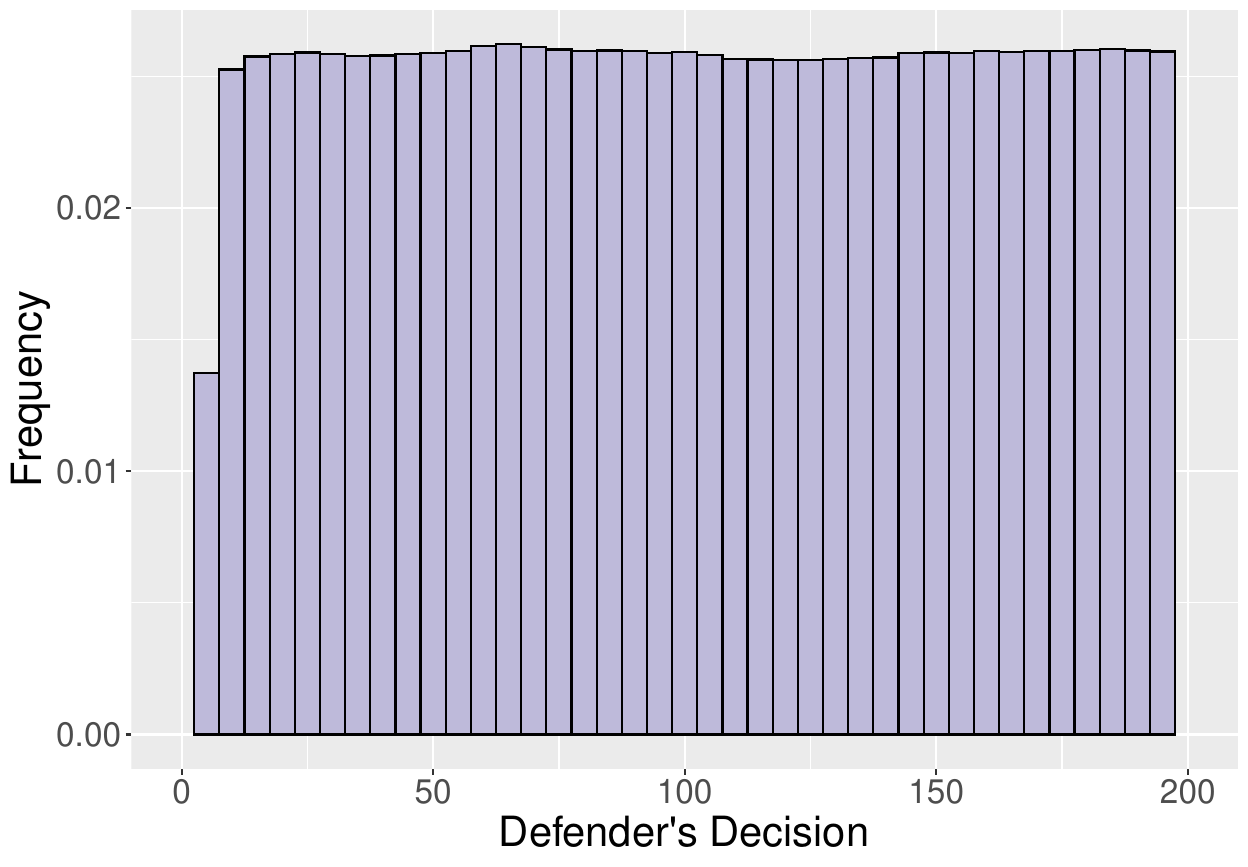}
\caption{APS solution of the Defender problem} \label{fig:apsprob3}
\end{subfigure}
\caption{ARA solution computed using APS}
\end{figure}

This is a problem with incomplete information and large decision spaces.
We compute the ARA solution using APS. We first estimate the probabilities $p_D(a \given d)$ for each defense $d$. %As in Section \ref{sec:computational_comp}, 
\textcolor{black}{
We replace $A$'s marginal augmented distribution $\pi_A$ by its power transformation $\pi^H_A$
to increase APS efficiency in finding the mode. In addition, as in simulated annealing, within each APS iteration we increase $H$ using an appropriate cooling schedule \citep{muller2004optimal}, see SM
Section 2.2 for details. Figure \ref{fig:pad} displays $p_D(a \given d)$ for four defenses $(0, 5, 10, 15)$. 
%{\color{black} 
If no defensive action is adopted ($d = 0$),
$A$ will launch the longest DDoS attack.
%}. 
Subscribing to a low protection plan ($d=5$) makes little practical difference. However, when increasing the protection to $10$ gbps, the attack forecast (from $D$'s perspective) becomes a mixture of high and low intensity values: small perturbations in $D$'s assumptions about $A$'s elements, induce big changes in the optimal attack. Finally, a $15$ gbps protection convinces $D$ that she should avoid the attack, attaining a deterrence effect.
The optimal solution remains the same for $d > 15$.}
%, and therefore results are not displayed.

%using Algorithm \ref{alg:ARA_APS_2}. %We model the uncertainty in the adversary's %parameter as in \cite{insua2019adversarial}.

%spread out across all possible alternatives. However, subscribing to a DDoS protection of just $50$ gbps, results in attacker alternatives to be concentrated in the region of lower values, attaining a deterrence effect. Increasing the protection to $195$ gbps makes no big difference.

\begin{figure}[htbp]
\begin{subfigure}{0.35\textwidth}
\centering
\includegraphics[width=\textwidth]{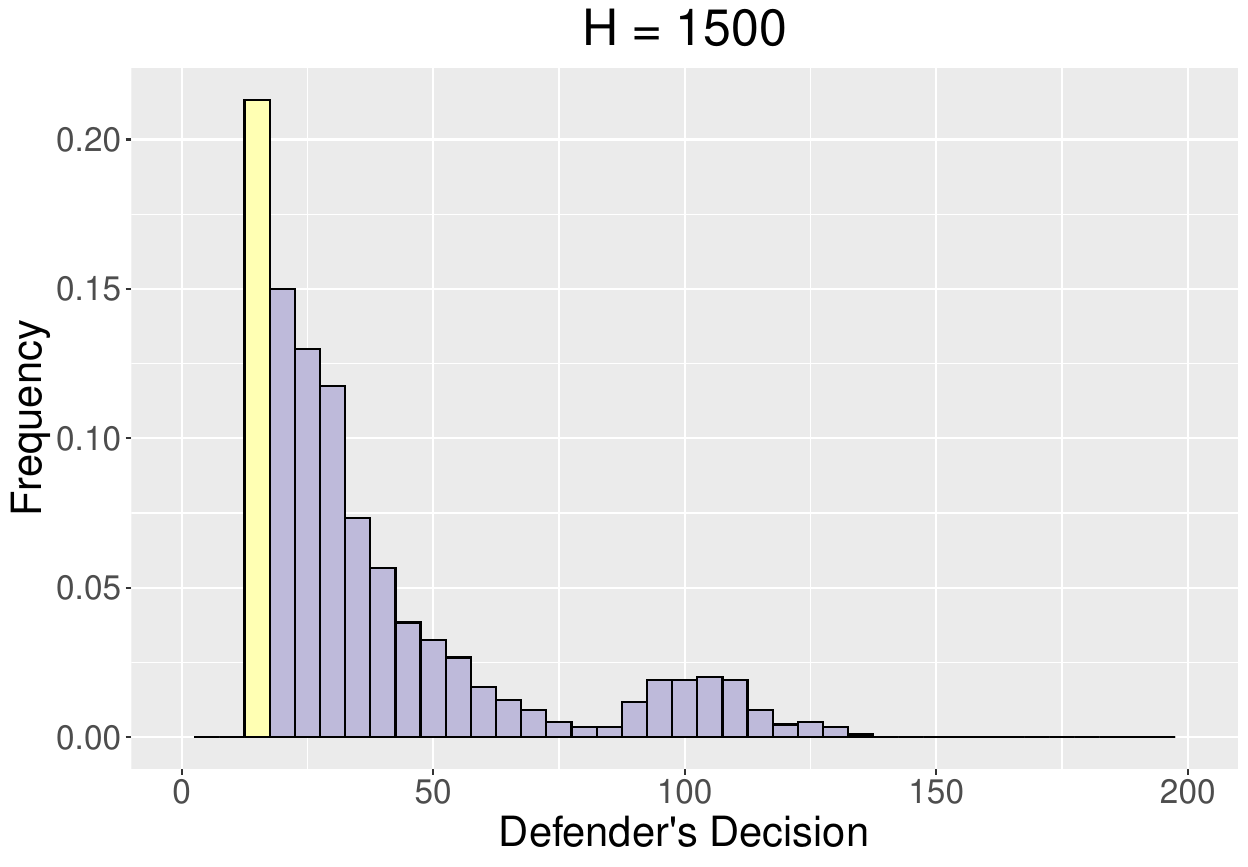}
\caption{}\label{fig:H1500}
\end{subfigure}
\hfill
\begin{subfigure}{0.35\textwidth}
\centering
\includegraphics[width=\textwidth]{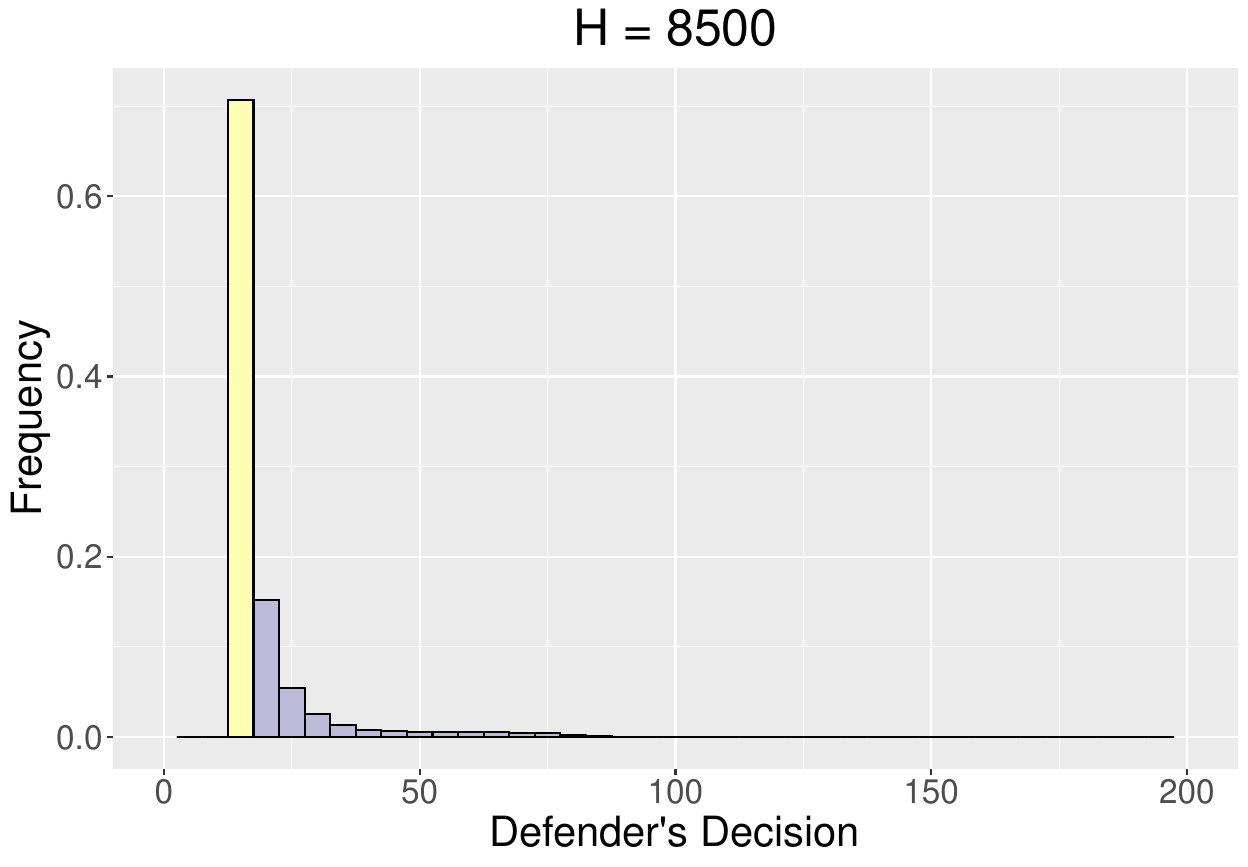}
\caption{}\label{fig:H8500}
\end{subfigure}

% \begin{subfigure}{0.4\textwidth}
% \centering
% \includegraphics[width=\textwidth]{aps_prob3_J4500.pdf}
% \caption{}\label{fig:H4500}
% \end{subfigure}
% \hfill
% \begin{subfigure}{0.4\textwidth}
% \centering
% \includegraphics[width=\textwidth]{aps_prob3_J8500.pdf}
% \caption{}\label{fig:H8500}
% \end{subfigure}
\caption{APS solutions of defender's problem for different augmentation parameter values}\label{fig:APS_Jtrick}%inverse temperature
\end{figure}

Figure \ref{fig:apsprob3} shows a histogram of the APS samples for $D$'s decision. As expected, the frequency of samples with value $15$ is similar to the ones with higher values. The histogram (and consequently the expected utility surface) is very flat. %, we cannot resolve the mode. 
\textcolor{black}{ Finding the exact optimal decision might not be crucial since the expected utilities for different protections are close. %That is a valid point. 
%the flat expected optimal utility region might be too big.
However, it should be noted that APS permits finding the optimal solution 
by sampling from a power transformation of the marginal augmented distribution,}
\textcolor{black}{
even with flat expected utilities, at a relatively small extra computational cost. This  
emphasizes another advantage of APS:} 
\textcolor{black}{ the provision of the distribution $\pi_D(d \given a^*(d))$ as part of the solution, delivering sensitivity analysis at no extra cost.} The decision maker may want to take such sensitivity of the optimal decision into account, and consider a defense which could be more robust. 
Thus, as with $A$'s problem, we sample iteratively, increasing the power $H$ of $D$'s marginal augmented distribution at each iteration.
 %For low values of $H$, Figure \ref{fig:H500}, convergence is clearly not achieved.
Figure \ref{fig:APS_Jtrick} illustrates how increasing $H$ makes the distribution more peaked around the optimal decision $d^*_{ARA} = 15$,
the cheapest plan that avoids the attack from $D$'s perspective.

%%%%%%%%%%%%%%%%%%%%%%%%%%%%%%%%%%
\section{Discussion}\label{sec:disc}

\textcolor{black}{
This paper considers providing robust support to a decision maker facing adversaries
in an environment where random consequences depend on the actions of all participants: 
%The prevalent paradigm under complete information is game-theoretic. When robustness assumptions do not hold, the proposed computational framework relaxes the common knowledge assumption: the
%resulting game with incomplete information is analyzed from the decision analytic perspective through ARA. 
\textcolor{black}{ we propose a framework that computes \textcolor{black}{a} game-theoretic solution under assumptions of complete information and perform a sensitivity analysis; if the solution is stable, it may be used with confidence and no further analysis is required.
Otherwise, or when complete information is actually lacking, we %relax the above assumption and 
use ARA as a decision analytic approach that mitigates common prior assumptions.} If such solution is stable, one may use it with confidence and stop the analysis. Otherwise, one must refine the relevant probability
and utility classes, eventually declaring the robustness of the ARA solution. If the solution is not robust, one could undertake a minimum regret or other (robust) analysis.}

We have provided MC and APS methods to support this process.
With large, discrete decision spaces, APS would be more efficient as its complexity does not depend on the decision sets' cardinalities. 
%{\color{black} 
In addition, %unlike MC, 
{\color{black}APS does not need discretization and can be directly applied to problems with continuous decision sets, for which MC may not provide feasible solutions at the desired precision.}
%} 
%We have provided empirical evidence supporting these ideas through our experiments. 
It should also be noted that MC errors associated with approximating the expected utility could overwhelm the calculation of the optimal decision. Samples from $p(\theta \given d, a)$ 
will typically need to be recomputed for each pair $(d,a)$. In contrast, APS performs the expectation and optimization simultaneously, sampling $d$ from regions with high utility with draws of $\theta $ from the utility-tilted 
augmented distribution. % rather than the conditional density $p(\theta \given d, a)$.
 %This results in draws of $\theta $ witmore frequently from higher utility. 
This reduces the MC error as the optimization effort in parts of the parameter space with low utility values is limited, resulting in reduced sample sizes for the same precision.
%{\color{red} In problems with continuous decision sets, an extension of the proposed approaches could use MC to limit the area of the decision space where the optimum is located, and then switching to APS to search within that area in more detail.} 
%In the continuous case, 
%as illustrated 
% in this setting. %\cite{2019arXiv190806901N}. 
Besides, when the expected utility surface is flat, MC simulation may need many draws or result in poor estimates, being also inefficient for random variables with skewed distributions. 
 %MC simulation may not be able to take that into account.
APS better handles those cases as it is based on sampling from the optimizing portions of the decision space; moreover, it can sample from a power transformation of the marginal augmented distribution, %less flat and 
more peaked around the mode. 
APS also provides sensitivity analysis at no extra computational cost. 

\textcolor{black}{ To keep the discussion concise, we illustrated the proposed framework 
with two stage (defend-attack) %sequential 
games, although it extends to \textcolor{black}{multi-stage sequential} games %. Indeed, the APS-based algorithmic approach can be extended to tackle multi-stage sequential games (
\textcolor{black}{with complete or incomplete information. }%). %for both in their classical and their ARA versions. 
For instance, for the sequential \textcolor{black}{d}efend-\textcolor{black}{a}ttack game with complete information, we proved that finding Nash equilibria is  equivalent to finding the mode of the marginal Defender's augmented distribution that uses the Attacker's APS output. This setup extends to solving $n$-stage sequential games by nesting $n$ APS algorithms (details in Section 3).}
\textcolor{black}{Similarly, the proposed framework is conceptually applicable to multi-stage games with incomplete information as well (Section 4 of the SM).
While extensions to $n$-stage games are technically viable, from a computational perspective they would suffer from the standard 
inefficiencies of dynamic programming. Thus, for a large number of stages, it may be worth exploring reinforcement learning based strategies \citep{powell2019unified}.
Overall, multi-stage sequential games would benefit from further illustration and computational experiments, which are left for future research. %Similarly, 
Moreover, extensions to simultaneous defend-attack games and the 
generic defend-attack interactions in the general bi-agent influence
diagrams in \cite{BAIDS} could be considered for %useful in 
security applications.}

Regarding algorithmic aspects, there is relevant future work. We have explored MC and APS as techniques to approximate Nash and ARA solutions in sequential two-stage games.
We highlight that although we have used APS with MH, other MCMC samplers could be used. %In particular, in problems with continuous decision sets in which gradient information about the utilities and probabilities is available, Hamiltonian Monte Carlo %(HMC) 
%techniques \citep{bishop2006pattern} could be applied.
Furthermore, {\color{black} as mentioned, MC and APS may not be the best choices for specific settings, and it would be worth exploring extensions of existing simulation optimization methods \citep{nelson2001simple}. \cite{naveiro2019gradient} propose a methodology to solve Stackelberg games with certain outcomes and continuous decision sets where gradient information about the utilities and probabilities is available. This work could be extended to games with uncertain outcomes,
%replacing the APS Metropolis-Hastings scheme 
by using Hamiltonian Monte Carlo (HMC) 
techniques \citep{bishop2006pattern} in place 
of our MH algorithms.} %to sample from the augmented distribution.} 
%For instance, in games with continuous and high dimensional decision sets, available gradient information could be exploited. %A limited version of this was used to solve Stackelberg games with certain outcomes \cite{naveiro2019gradient}, however 
%When the gradient cannot be directly evaluated, stochastic approximation or response surface methods \citep{fu2015handbook} could be potential alternatives. %In addition, using APS in combination with graphical methods to simplify Bayesian games \citep{thwaites2018graphical} is a fruitful line of research.

%Note that we have restricted to cases in which the Attacker's best response is unique. When this is not the case, we could straightforwardly extend our algorithms to consider variants of the sequential defend-attack game in which the adversary breaks ties so that the response would favor or hurt the defender the most.

%It is also worth mentioning that, as mentioned, i
In the last  decade, game-theoretic models have gained importance in AML \citep{AMLARA}. Efficient scalable algorithmic approaches to solve typical games appearing in this new context are essential. In addition, the ARA solution of such games turn to be important, as common 
knowledge assumptions rarely hold in AML.
Thus, this paper would contribute to that end, proposing a novel simulation based approach to solve sequential defend-attack games both from the classic and the ARA perspectives.
%}

%From a computational complexity perspective, note that
%Monte Carlo simulation takes  $D*N*A + D*N$ steps in addition to the cost of optimization, whereas the
%standard and nested APS approaches, respectively, require $N * (D+A)$ and $N*(G+1)*(D+A) $ steps in addition to mode computation and convergence checks.
% Therefore, APS  can be a preferred choice in cases of many decision alternatives since its computational burden grows linearly rather than geometrically.
%The standard implementation of APS can be a good alternative in case of relative low dimensionality of number of stages (and alternatives). When the dimensionality %increases, the computation can be cumbersome, and one can use the nested algorithm. There is a trade-off between the cost of mode computation and increased accuracy.

{\small 
\paragraph{Acknowledgements}{T.E. acknowledges the support of Texas State University through Steven R. “Steve” Gregg Endowed Professorship
%, Faculty Development Leave
and Presidential Research Leave Award. R.N. acknowledges support of the Spanish Ministry of Education for his FPU15-03636 grant. D.R.I. is grateful to the MTM2017-86875-C3-1-R AEI/FEDER EU project, the AXA-ICMAT Chair in ARA and the EU's Horizon 2020 project 740920 CYBECO. This material is based upon work supported by the Air Force Scientific Office of Research (AFOSR)  award FA-9550-21-1-0239, AFOSR European Office of Aerospace Research and Development award FA8655-21-1-7042 and was partially supported by the National Science Foundation under Grant DMS-1638521 to the SAMSI, %Statistical and Applied Mathematical Sciences Institute, 
and the AMALFI BBVA Foundation project.% and a European Office of Aerospace Research and Development %(EOARD) award.  
}
}



%\begin{thebibliography}{00}


%\bibitem{}

%\end{thebibliography}

%%%%%%%%%%%%%%%%%%%%%%%%%%%%

%%%%%%%%%%%%%%%%%%%%%%%%%%%%%%%%%%%%%%%%%%%%%%%%%%%%%%%%%%%%%%%%%%%%

\bibliography{aps_biblio}

%\clearpage

\glsaddall
\printglossary[type=\acronymtype,style=mcolindex, title=Acronyms]

\appendix
\section{Proofs of Propositions}\label{apdx:proofs}

\noindent \textit{Proposition \ref{prop:conv_nested}}: 
%\textcolor{red}{First, we show the optimality of attacker's decision, and use that as input for the defender's decision problem to to compute the optimal game theoretic solution under complete information.}
For a fixed $d$, given the assumptions about $u_A$ and $p_A$,
$\psi_A (a,d)$ is continuous in $a$ and, since 
$\mathcal{A}$ is compact, there exists 
$a^* (d)= \argmax _{a\in \cal A} \psi _A (a,d)$, which is continuous in $d$.
%(and unique for each $d$ by assumption)%\textcolor{red}{For simplicity, assume that the best response $a^* (d)$ is unique.}
\textcolor{black}{ Moreover, $\pi_A(a, \theta \given d)\propto u_A (a, \theta) p_A (\theta | d,a) $ is well-defined. 
%and $\pi_A(a \given d)$ has a unique global mode for each $d$.
Using \cite{smith1993bayesian} convergence results 
for MH algorithms,
the samples generated by the inner APS loop in Alg.\ref{alg:MHdefenderAPS2} define a %MH 
Markov chain with  $\pi_A(a, \theta \given d)$ as stationary distribution. 
%, since the support of  
%$g_A \times p_A $ coincides with that of 
%the target distribution $\pi_A$. 
Once convergence is detected, %(at iteration $K$), 
the 
%for each $d$,
%. Once MCMC convergence is assessed, the samples of $\theta$ are discarded since the samples of $a$ are of interest. The first $K$ samples of $a$ are discarded as burn-in samples. The 
remaining $M-K$ marginal samples $a^{(i)}$ of the chain are approximate samples
from $\pi _A (a | d)$. % which, recall, was 
%proportional to $\psi_A (a, d)$. 
Then, for a large enough $M-K$, 
a consistent sample mode estimator}, \textcolor{black}{ in the sense of \cite{chen},
 based on such approximate 
sample detects  $a^\opt (d)$ a.s., from which we select 
the desired mode.}
%Using the strong law of large numbers, such almost sure convergence for the discrete case is written as: $\argmax \lbrace\# a; a \in \lbrace a_{(K+1)},...,a_{(M)}  \rbrace \rbrace \rightarrow mode[\pi_A(a \given d)]= a^{*}(d)$. 
%For simplicity, we assume that the best response $a^* (d)$ is unique. In cases where the Attacker's optimal response is not unique, we could consider variants of the Stackelberg Games, where the adversary could break the ties so that the response would favor or hurt the defender the most.

Next, under the stated conditions, and taking 
into account the continuity of $a^*(d)$ in $d$,
$\psi_D (d, a^*(d))$ is continuous in $d$.
Since ${\cal D}$ is compact, there exists $d_{GT}^*$.
Besides, $\pi_D(d, \theta \given a^\opt(d))
\propto u_D (d, \theta ) p_ D (\theta | d, a^*(d))$ 
is well-defined. %By repeating the  argument of the inner APS, t
 The samples generated
by the outer APS in Algorithm \ref{alg:MHdefenderAPS2} define a 
Markov chain
with  $\pi_D(d, \theta \given a^\opt(d))$ 
as stationary distribution.
%ince the domain of $g_D$ coincides with that of
%$\pi _D$, using again the results in 
%Roberts and Smith (1993). 
%Then,  
Once convergence is detected, %(at iteration $R$), 
%d since the samples of $d$ are of interest. The first $R$ samples of $d$ are discarded as burn-in samples. The
the remaining $N-R$ marginal samples $d^{(i)}$ 
constitute an approximate sample
from $\pi_D(d \given a^\opt(d))$. %Based on it,
Then, a consistent sample mode estimator a.s.
convergent to $d^\opt_\text{GT}$ 
\textcolor{black}{ may be built \citep{chen}, by
selecting the appropriate mode, facilitated, if wished, by 
the peaking trick. \hfill $\triangle$}

%END \section{Appendix} \ref{apdx:proofs}

\noindent \textit{Proposition \ref{conv_apsara}}: 
%\textcolor{red}{First, we show the optimality of attacker's decision, and use that as input for the defender's decision problem to to compute the optimal ARA solution.}
%
\textcolor{black}{ For each $d$ and $\omega\in \Omega$, given 
the assumptions about $U_A(a, \theta)$,  
$P_A(\theta \given d,a) $ and ${\cal A}$, we have:
1) $\Psi ^\omega _A (d, a) =\int U^{\omega }_A (a, \theta ) P_A ^{\omega} (\theta | d, a)$ is a.s.\ continuous in $a$;
2) there exists $A (d)^{*\omega } = \argmax 
\Psi ^\omega _A (d, a)$ a.s.\
    which is continuous in $d$; 
%(\textcolor{red}{, where again for simplicity, we assume that the best response is unique.}
3) the distribution 
$ \Pi_A^{\omega} (a, \theta \given d)$ 
 $\propto U_A^{\omega }(a, \theta)P_A^{\omega }(\theta \given d,a)$
is well defined.}
The samples generated through the  \texttt{sample\_attack} function are distributed according to $p_D(a|d)$. By construction, through sampling $u_A(a, \theta) \sim U_A(a, \theta)$ and $p_A(\theta \given d, a) \sim P_A(\theta \given d, a)$, one can build $\pi_A(a, \theta \given d) \propto u_A(a, \theta) p_A(\theta \given d, a)$, which is a sample from $\Pi _A (a, \theta \given d)$. %Then, $\text{mode}(\pi_A(a \given d))$ is a sample of $A^*(d)$, whose distribution is $\Pbb_F \left[ A^* (d) = a  \right] = p_D(a \given d)$. 
Following an argument similar to that of Proposition 1,
the $a$ samples generated through the MH routine in \texttt{sample\_attack} 
%in Algorithm \ref{alg:ARA_APS_2} 
define a Markov chain whose stationary 
distribution is %$\pi_A(a, \theta \given d)$
 $ \pi_A(a, \theta \given d) \sim \Pi _A (a, \theta \given d)$. %In addition, $\pi_A(a, \theta \given d)$ has a unique global mode, by assumption.
Once convergence is detected, \textcolor{black}{ a consistent estimator of the modes of the $a$ samples can be built
which converges a.s.\ to $\text{mode}(\pi_A(a \given d))$, whose distribution is $\Pbb_F \left[ A^* (d) \leq a  \right] = p_D(A\leq a \given d)$.%, by selecting the appropriate mode.
}

%Indeed, as we are sampling $u_A(a, \theta) \sim U_A^{\omega}(a, \theta)$ and $p_A(\theta \given d, a) \sim P_A^{\omega}(\theta \given d, a)$; $\pi_A(a, \theta \given d) \propto u_A(a, \theta) p_A(\theta \given d, a)$, is a sample from $\Pi_A^{\omega}(a, \theta \given d)$. Then, $\text{mode}(\pi_A(a \given d))$ is a sample from $A^*(d)^{\omega}$, whose distribution is $\Pbb_F \left[ A^* (d) = a  \right]$. To compute this mode, we sample $a, \theta \sim \pi_A(a, \theta \given d)$ using MH, defining a Markov chain $\lbrace a^{(i)}, \theta^{(i)}; i=1,...,M \rbrace$ (loop 
%in function \texttt{sample\_attack}) whose stationary distribution
%is $\pi_A(a, \theta \given d)$, \citep{roberts1994simple}.
%Once MCMC convergence is assessed, the
%first $K$ samples of $a$ are discarded as burn-in 
%and the remaining $M-K$ marginal samples are approximate samples
%from $\pi_A(a \given d)$. For large enough $M-K$, the mode of such 
% marginal samples approximates the mode of $\pi_A (a, \theta \given d)$,
% based on a consistent mode estimator \citep{romano1988weak}.
%
As ${\cal D}$ is compact 
and $\psi_D (d)$ is continuous,  $d^\opt_\text{ARA}$ exists.
Since  $u_D$ is positive and integrable, 
$\pi_D(d,a,\theta) \propto u_D(d, \theta)p_D(\theta \vert d,a)p_D(a \vert d)$ 
is  
well-defined,  %Following the arguments in Proposition 1, this 
and is the stationary distribution of the MH Markov chain of the outer APS routine in Algorithm \ref{alg:ARA_APS_2}.
Once convergence is detected, %the samples of $\theta$ are discarded since the samples of $d$ are of interest. T
 the marginal samples in $d$ are approximately distributed as $\pi_D(d)$,
 from which we approximate a.s.
$d^\opt_\text{ARA}$ through a \textcolor{black}{consistent mode 
estimator. \hfill $\triangle$
%,
% as in \cite{chen}, 
%similar to above. %, and choosing the appropriate mode, possibly using the peaking trick}.
}

%%%%%%%%%%%%%%%%%%%%%%%%%%%%%%%%%%%

\end{document}